\documentclass[12pt]{article}
\usepackage{amsmath}
\usepackage{graphicx}
\usepackage{bm}
\usepackage{fancyhdr}
\usepackage{amssymb}
\usepackage{setspace}
\usepackage{color}
\usepackage{epsfig}
\usepackage{overpic}
\usepackage{caption,subcaption}
\usepackage{epstopdf}
\usepackage{ulem}
\usepackage{cancel}
\begin{document}

\baselineskip=20pt

\newcommand{\Title}[1]{{\baselineskip=26pt
   \begin{center} \Large \bf #1 \\ \ \\ \end{center}}}
\newcommand{\Author}{\begin{center}
   \large \bf
Wei Wang ${}^{a,b}$, Yi Qiao ${}^{b}$, Rong-Hua Liu ${}^{a}$, Wu-Ming Liu${}^{b,c,d}\footnote{Corresponding author: wmliu@iphy.ac.cn}$ and Junpeng Cao ${}^{b,c,d,e}\footnote{Corresponding author: junpengcao@iphy.ac.cn}$
 \end{center}}
\newcommand{\Address}{\begin{center}
     ${}^a$ National Laboratory of Solid State Microstructures, School of Physics and Collaborative Innovation Center of Advanced
           Microstructures, Nanjing University, Nanjing 210093, China\\
     ${}^b$ Beijing National Laboratory for Condensed Matter Physics, Institute of Physics, Chinese Academy of Sciences, Beijing
           100190, China\\
     ${}^c$ School of Physical Sciences, University of Chinese Academy of Sciences, Beijing 100049, China\\
     ${}^d$ Songshan Lake Materials Laboratory, Dongguan, Guangdong 523808, China\\
     ${}^e$ Peng Huanwu Center for Fundamental Theory, Xian 710127, China\\
   \end{center}}

\Title{Elementary excitations in an integrable twisted $J_1-J_2$ spin chain in the thermodynamic limit} \Author

\Address \vspace{0.1cm}

\vspace{1truecm}

\begin{abstract}

The exact elementary excitations in a typical $U(1)$ symmetry broken quantum integrable system, that is
the twisted $J_1-J_2$ spin chain with nearest-neighbor, next nearest neighbor and chiral three spin interactions,
are studied. The main technique is that we quantify the energy spectrum of the system by the
zero roots of the transfer matrix instead of the traditional Bethe roots.
From the numerical calculation and singularity analysis, we obtain the patterns of zero roots.
Based on them, we analytically obtain the ground state energy and the elementary excitations in the thermodynamic limit.
We find that the system also exhibits the nearly degenerate states in the regime of $\eta\in \mathbb{R}$,
where the nearest-neighbor couplings among the $z$-direction are ferromagnetic.
More careful study shows that the competing of interactions can induce the gapless low-lying excitations and quantum phase transition in the
antiferromagnetic regime with $\eta\in \mathbb{R}+i\pi$.

\vspace{1truecm}

%75.10.Pq Spin chain models
%02.30.Ik Integrable systems
%71.10.Pm Fermions in reduced dimensions (anyons, composite fermions, Luttinger liquid, etc.

\end{abstract}

\newpage

%%%%%%%%%%%%%%%%%%%%%%%%%%%%%%%%%%%%%%%%%%%%%%%%%%%%%%%%%%%%%%%
%                                                             %
%  1. Introduction                                            %
%                                                             %
%%%%%%%%%%%%%%%%%%%%%%%%%%%%%%%%%%%%%%%%%%%%%%%%%%%%%%%%%%%%%%%
\hbadness=10000

\tolerance=10000

\hfuzz=150pt

\vfuzz=150pt
\section{Introduction}
\setcounter{equation}{0}

Understanding the collective behavior in the one-dimensional quantum many-body systems is a fascinating and challenging issue.
Due to the competition of some kinds of interactions, many novel physical phenomena are found, and new physical pictures are developed.
The exact solution can provide the benchmark of these new theories \cite{Onsa, Lieb,Car18}.
The typical methods of seeking the exact solution are the coordinate \cite{Beth31} and algebraic Bethe ansatz \cite{Tak79,Fad80,Skl80,Skl82,Ku82,Tak85,Skl88}, as well as the $T-Q$ relation \cite{Bax71,Bax712}.
These methods are powerful when studying quantum integrable systems with $U(1)$ symmetry.
However, if the $U(1)$ symmetry is broken, it is hard to construct a suitable reference state
and to apply these methods. On the other hand, based on the Yang-Baxter equation and
reflection equations, we can prove that there indeed exist some quantum integrable systems without $U(1)$ symmetry.
The next problem is how to solve them exactly. Then many interesting methods such as
gauge transformation \cite{cao03}, $T-Q$ relation based on the fusion \cite{Yung95,nep021}, $q$-Onsager algebra \cite{Bas1,Bas2},
separation of variables \cite{sk2-2,Niccoli13-1}, modified algebraic Bethe ansatz \cite{Bel13-1,Bel13-2,Bel13-3}
and off-diagonal Bethe ansatz \cite{cysw,Book} have been developed.
We should note that the exact solutions of quantum integrable systems without $U(1)$ symmetry have many applications
in non-equilibrium statistical mechanics \cite{Gier05,Sir09}, topological physics \cite{Fu08,Lut10} and high energy physics \cite{Mal98,Bei12,Jiang20,Leeuw21,Ber05}.

The next question is how to calculate the exact physical quantities of the systems in the thermodynamic limit.
The difficulties come from the eigenvalues, and the associated Bethe ansatz equations (BAEs) are inhomogeneous.
Thus it is impossible to take the logarithm of BAEs and use the thermodynamic Bethe ansatz.
Recently, a novel Bethe ansatz scheme has been proposed to calculate the physical quantities of quantum integrable systems with or without $U(1)$ symmetry to overcome the obstacles \cite{Qiao21,Le21}.
The main idea is that the eigenvalues of the transfer matrix can be characterized by their zero roots instead of the traditional Bethe roots.

In this paper, we study an integrable $J_1-J_2$ spin chain which includes the nearest-neighbor (NN), next-nearest-neighbor (NNN) and chiral three-spin interactions.
The boundary condition is the antiperiodic one. The twisted boundaries break the $U(1)$ symmetry of the system.
After the boundary reflection, the spins of quasi-particles are not conserved.
Based on the algebraic analysis, we obtain the energy spectrum of the system and the homogeneous BAEs.
From the numerical calculation and singularity analysis of BAEs, we get the distributions of solutions in the thermodynamic limit.
Then we compute the ground state energy and elementary excitations.
We also find the nearly degenerate states in the ferromagnetic regime and the quantum phase transition in the antiferromagnetic regime.

The paper is organized as follows. The next section serves as an introduction to the antiperiodic $J_1-J_2$ spin chain and the explanation of its integrability.
In section 3, we give the eigenvalues spectrum.
In section 4, combined with the inhomogeneous $T-Q$ relation, we analyse the zero root patterns of the eigenvalue of the transfer matrix.
We study the nearly degenerate states in section 5.
Then, we calculate the ground state energy and low-lying excitations in the thermodynamic limit focusing on the regime of real in section 6.
In section 7, the exact physical properties are discussed in the regime of $\eta\in \mathbb{R}+i\pi$.
Concluding remarks and discussions are given in section 8.

\section{The system and integrability}
\setcounter{equation}{0}

\begin{figure}[t]
\begin{center}
\includegraphics[width=8cm]{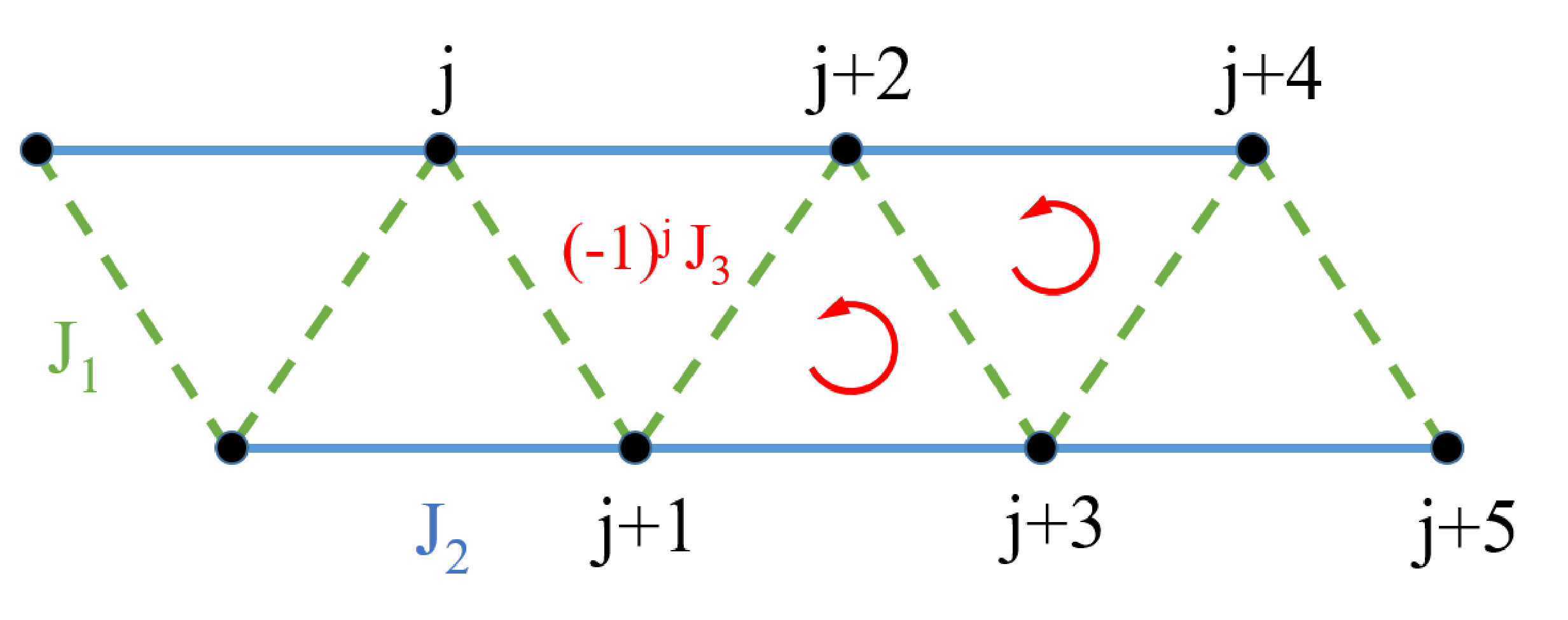}
\caption{Illustration of the bulk of twisted $J_1-J_2$ spin system with the nearest-neighbor interactions $J_1$, next-nearest-neighbor interactions $J_2$ and chiral three-spin interactions $(-1)^jJ_3$. The model can be viewed as the zigzag chain. The order of the chiral three-spin coupling in each triangle is indicated by the red arrows.}\label{J1J2model}
\end{center}
\end{figure}

The Hamiltonian of the integrable anisotropic $J_1-J_2$ model reads
\begin{eqnarray}\label{Ham1}
H =-\sum^{2N}_{j=1} \sum_{\alpha=x,y,z}  \big[ J_1^\alpha \sigma_j^\alpha \sigma_{j+1}^\alpha
+J_2 \sigma_j^\alpha \sigma_{j+2}^\alpha + (-1)^j J_3^{\alpha} \sigma_{j+1}^\alpha(\vec{\sigma}_{j}  \times \vec{\sigma}_{j+2} )^\alpha\big],
\end{eqnarray}
as shown in Fig.\ref{J1J2model}.
Here $2N$ is the number of sites.
$\{\sigma^{\alpha}_j|j=1,\cdots, 2N\}$ are the $2\times2$ Pauli matrices along the $\alpha$-direction at the $j$-th site.
In this paper, we consider the system \eqref{Ham1} with antiperiodic boundary condition
\begin{eqnarray}
\sigma^{\alpha}_{2N+n}=\sigma^{x}_{n} \sigma^{\alpha}_{n} \sigma^{x}_{n},\quad  n=1,2, \quad \alpha=x,y,z,\label{APB}
\end{eqnarray}
which gives $\sigma^{x}_{2N+n}=\sigma^{x}_{n}$, $\sigma^{y}_{2N+n}=-\sigma^{y}_{n}$ and $\sigma^{z}_{2N+n}=-\sigma^{z}_{n}$.
$J_1^\alpha$ quantifies the NN coupling with the form of
\begin{eqnarray}
J_1^x=J_1^y=\cosh(2a), \quad J_1^z=\cosh\eta,
\end{eqnarray}
where $a$ is the model parameter and $\eta$ is the anisotropic parameter. $J_2$ characterizes the NNN isotropic coupling,
\begin{eqnarray}
J_2=-\frac{\sinh^2(2a)\cosh\eta}{2\sinh^2\eta}.
\end{eqnarray}
$J^\alpha_3$ describes the chiral three-spin coupling with the strength of
\begin{eqnarray}
J_3^x=J_3^y= \frac{i \sinh(2a)}{2 \sinh\eta}\cosh\eta, \quad J_3^z=\frac{i \sinh(4a)}{4 \sinh\eta}.
\end{eqnarray}
If $a=0$, the model \eqref{Ham1} degenerates into the Heisenberg spin chain.
It is worth mentioning that the hermitian of Hamiltonian (\ref{Ham1}) requires that $a$ must be real if $\eta$ is imaginary, and $a$ must be imaginary if $\eta$ is real or $\eta\in \mathbb{R}+i\pi$.

Now, we show that the model (\ref{Ham1}) is integrable, which is related to the six-vertex $R$-matrix
\begin{eqnarray}
R_{0,j}(u)=\frac{\sinh(u+\eta)+\sinh u}{2\sinh \eta}+\frac{1}{2} (\sigma^x_j \sigma^x_0 +\sigma^y_j \sigma^y_0)
+ \frac{\sinh(u+\eta)-\sinh u}{2\sinh \eta} \sigma^z_j \sigma^z_0,
\label{R-matrix}
\end{eqnarray}
where $u$ is the spectral parameter.
Throughout this paper, we adopt the standard notations. For
any matrix $A\in {\rm End}(\mathbb{C})$, $A_j$ is an embedding operator
in the tensor space $\mathbb{C}^2\otimes \mathbb{C}^2\otimes\cdots$, which acts
as $A$ on the $j$-th space and as identity on the other factor
spaces. $R_{0,j}(u)$ is an embedding operator of $R$-matrix in the
tensor space, which acts as identity on the factor spaces except
for the $0$-th and $j$-th ones. Here 0 means the auxiliary space and $j=1,\cdots,2N$ means the physical or quantum space.
The $R$-matrix (\ref{R-matrix}) has the following  properties
\begin{eqnarray}
&&\hspace{-1.5cm}\mbox{ Initial
condition}:\,R_{0,j}(0)= P_{0,j},\label{Int-R} \\
&&\hspace{-1.5cm}\mbox{ Unitarity
relation}:\,R_{0,j}(u)R_{j,\,0}(-u)= \phi(u)\times{\bf id}, \label{Unitarity} \\
&&\hspace{-1.5cm}\mbox{ Crossing
relation}:\,R_{0,j}(u)=V_0R_{0,j}^{t_j}(-u-\eta)V_0,\quad V_0=-i\sigma_0^y,
\label{crosing-unitarity} \\
&&\hspace{-1.5cm}\mbox{ PT-symmetry}:\,R_{0,j}(u)=R_{j,\,0}(u)=R^{t_0\,t_j}_{0,j}(u),\label{PT} \\
&&\hspace{-1.4cm}\mbox{$Z_2$-symmetry}: \;\;
\sigma^\alpha_0\sigma^\alpha_jR_{0,j}(u)=R_{0,j}(u)\sigma^\alpha_0\sigma^\alpha_j,\quad
\mbox{for}\,\,\,
\alpha=x,y,z,\label{Z2-sym} \\
&&\hspace{-1.5cm}\mbox{ Quasi-periodicity}:\, R_{0,j}(u+i\pi)=-\sigma^z_0R_{0,j}(u)\sigma^z_0,\label{quasi-}\\
&&\hspace{-1.5cm}\mbox{ Fusion relation}:\, R_{0,j}(-\eta)=-2P^{(-)}_{0,j}.\label{fu-}
\end{eqnarray}
Here $\phi(u)=-\sinh(u+\eta)\sinh(u-\eta)/\sinh^2\eta$, ${\bf id}$ is the identity operator, $R_{j,0}(u)=P_{0,j}R_{0,j}(u)P_{0,j}$ with $P_{0,j}$ being
the permutation operator, $t_l$ denotes transposition in
the $l$-th space, and $P^{(-)}_{0,j}$ is the one-dimensional antisymmetric projection operator, $P^{(-)}_{0,j}=(1- P_{0,j})/2$. Besides, the $R$-matrix (\ref{R-matrix}) satisfies the Yang-Baxter equation
\begin{eqnarray}
R_{0,j}(u_1-u_2)R_{0,l}(u_1-u_3)R_{j,l}(u_2-u_3)
=R_{j,l}(u_2-u_3)R_{0,l}(u_1-u_3)R_{0,j}(u_1-u_2).\label{QYB}
\end{eqnarray}

We combine all the $R$-matrices in different sites and define the monodromy matrix as
\begin{eqnarray}\label{monodromy-matrix}
T_0(u)=\sigma^x_0 R_{0,1}(u-\theta_1) R_{0,2}(u-\theta_2) \cdots R_{0,2N-1}(u-\theta_{2N-1}) R_{0,2N}(u-\theta_{2N}),
\end{eqnarray}
where the $\{\theta_j| j=1,\cdots, 2N\}$ are the inhomogeneous parameters.
The transfer matrix is given by tracing the monodromy matrix in the auxiliary space
\begin{eqnarray} \label{trans}
t(u)=tr_0 T_0(u).
\end{eqnarray}
Based on the commutation relation (\ref{Z2-sym}) and the Yang-Baxter equation (\ref{QYB}),
one can prove that the transfer matrices with different spectral parameters commutate with each other, i.e.,
\begin{eqnarray}\label{t-commu1}
[t(u), t(v)]=0.
\end{eqnarray}
Expanding the transfer matrix with respect to the spectral parameter $u$, all the expansion coefficients are also commutative.
According to the quantum integrable theory, all the conserved quantities including the model Hamiltonian
can be constructed by these expansion coefficients. Because the number of independent conserved quantities equals to that of the degrees of freedom,
the system is integrable.

The Hamiltonian (\ref{Ham1}) is generated by the transfer matrices as
\begin{eqnarray}\label{J1J2ham}
H = \phi^{1-N}(2a)\sinh\eta\big\{ t(a-\eta)\frac{\partial \, t(u)}{\partial u}\big|_{u=a}
  + t(-a-\eta) \frac{\partial \, t(u)}{\partial u}\big|_{u=-a} \big\}\big|_{\{\theta_j=(-1)^ja\}}+E_0,
\end{eqnarray}
where the constant $E_0$ is given by
\begin{eqnarray}
E_0=-\frac{N\cosh\eta[\cosh^2(2a)-\cosh(2\eta)]}{\sinh^2\eta}.
\end{eqnarray}

Another interesting conserved quantity is the shift operator, which is generated by the transfer matrix as \cite{Wang22}
\begin{eqnarray}
U=\phi^{-N}(2a)t(a)t(-a)|_{\{\theta_j=(-1)^ja\}}.
\end{eqnarray}
One can find that the operator $U$ commutate with the Hamiltonian.
The $U$ characterizes the transition invariance of the present system. According to the quantum theory, we can define
the topological momentum $k$ as
\begin{eqnarray}\label{kz}
k=-i\ln U=-i\sum_{j=1}^{2N-1}\ln\frac{\sinh(a+z_j-\frac\eta2)}{\sinh(a-z_j-\frac\eta2)}{~~}mod\,\{2\pi\}.
\end{eqnarray}

\section{The eigenvalues spectrum}
\setcounter{equation}{0}

From the construction (\ref{J1J2ham}), we know that the eigen-energies of the Hamiltonian (\ref{Ham1}) are
related with the eigenvalues of transfer matrix $t(u)$. If the eigenvalues of the transfer matrix are known, the eigen-energies are known.
Thus we diagonalize the transfer matrix $t(u)$ first. Because the $U(1)$ symmetry of the system is broken,
we calculate the eigenvalues of transfer matrix based on the polynomial analysis.
The main ideas are as follows. From the definitions (\ref{R-matrix}), (\ref{monodromy-matrix}) and
(\ref{trans}), we know that the transfer matrix $t(u)$ is an operator-valued trigonometric polynomial with degree $2N-1$ due to the existence of twisted matrix $\sigma_0^x$ and partial trace.
According to the algebra analysis theory, the values of $t(u)$ can be completely determined by the $2N-1$ independent constraints.
Then we should seek these constraints, which can be achieved by using the fusion technique.
At the inhomogeneous point $\theta_j$, the $R$-matrix (\ref{R-matrix}) degenerates into the one-dimensional projector operator, please see Eq.(\ref{fu-}).
By using the properties of projector and Yang-Baxter equation, we can obtain the transfer matrices product identities in these one-dimensional subspaces
characterized by $\{\theta_j|j=1, \cdots, 2N\}$.
At the certain values of spectral parameter, these identities are closed
and can be used to determine the eigenvalues of $t(u)$. The more detailed description can be found in the reference \cite{cysw} or chapter 4 in the book \cite{Book}.

By using the initial condition (\ref{Int-R}), we have
\begin{eqnarray}
t(\theta_j)&=&tr_0\{ \sigma^x_0 R_{0,1}(\theta_j-\theta_1)\cdots P_{0,j} \cdots R_{0,2N-1}(\theta_j-\theta_{2N-1}) R_{0,2N}(\theta_j-\theta_{2N})\}\nonumber\\
          &=& R_{j,j+1}(\theta_j-\theta_{j+1})\cdots R_{j,2N}(\theta_j-\theta_{2N})\sigma^x_j R_{j,1}(\theta_j-\theta_1) \cdots R_{j,j-1}(\theta_j-\theta_{j-1}).\label{1}
\end{eqnarray}
with the help of crossing relation (\ref{crosing-unitarity}), the transfer matrix $t(\theta_j-\eta)$ can be calculated as
\begin{eqnarray}
t(\theta_j-\eta)&=&(-1)^{2N-1} R_{j,j-1}(-\theta_j+\theta_{j-1})\cdots R_{j,1}(-\theta_j+\theta_{1})\nonumber\\
          &&\times\sigma^x_j R_{j,2N}(-\theta_j+\theta_{2N}) \cdots R_{j,j+1}(-\theta_j+\theta_{j+1}).\label{2}
\end{eqnarray}
Multiplying Eqs.(\ref{1}) and (\ref{2}), and using the unitarity relation (\ref{Unitarity}), we obtain
following operators product identities
\begin{eqnarray}
t(\theta_j)t(\theta_j-\eta)=-a(\theta_j)d(\theta_j-\eta)\times{\bf id}, \quad j=1, \cdots, 2N, \label{theta}
\end{eqnarray}
where
\begin{eqnarray}
d(u)=a(u-\eta)=\prod_{j=1}^{2N}\,\frac{\sinh(u-\theta_j)}{\sinh\eta}.\label{ai1}
\end{eqnarray}

Denote the eigenvalue of $t(u)$ as $\Lambda(u)$. Acting the operator identities \eqref{theta} on a common state of $t(u)$ and $t(u-\eta)$, we obtain
following functional relations
\begin{eqnarray}
\Lambda(\theta_j)\Lambda(\theta_j-\eta)=-a(\theta_j)d(\theta_j-\eta), \quad j=1, \cdots, 2N.\label{theta1}
\end{eqnarray}
The eigenvalue $\Lambda(u)$ is a trigonometric polynomial of $u$ with the degree $2N-1$. Thus the value of
$\Lambda(u)$ can be completely determined by the $2N$ constraints \eqref{theta1}.

Besides, the transfer matrix $t(u)$ satisfies the periodicity
\begin{eqnarray}
t(u+i\pi)=(-1)^{2N-1}t(u),
\end{eqnarray}
which gives
\begin{eqnarray}
\Lambda(u+i\pi)=(-1)^{2N-1}\Lambda(u).\label{periodic}
\end{eqnarray}
According to Eqs.\eqref{theta1} and \eqref{periodic}, we express $\Lambda(u)$ in terms of its $2N-1$ zero roots $\{z_j-\eta/2|j=1,\cdots,2N-1\}$ and an overall coefficient $\Lambda_0$
as
\begin{eqnarray}
\Lambda(u)=\Lambda_0\,\prod_{j=1}^{2N-1}\,\sinh (u-z_j+\frac\eta2).\label{Zero-points}
\end{eqnarray}
Substituting the parameterization \eqref{Zero-points} into (\ref{theta1}), we obtain the constraints among zero roots
\begin{eqnarray}
&&\Lambda_0^2\prod_{j=1}^{2N-1}\sinh(\theta_l-z_j+\frac\eta2)\sinh(\theta_l-z_j-\frac\eta2)
=-\sinh^{-4N}\eta\,\prod_{j=1}^{2N}\sinh(\theta_l-\theta_j+\eta)\nonumber\\
&&\times \sinh(\theta_l-\theta_j-\eta),\quad l=1,\cdots,2N.\label{theta2}
\end{eqnarray}
We note the BAEs \eqref{theta2} are homogeneous.
From the construction (\ref{J1J2ham}), we obtain the energy spectrum of Hamiltonian (\ref{Ham1}) as
\begin{eqnarray}\label{J1J2redu}
E&=&\phi^{1-N}(2a)\sinh\eta \big\{ \Lambda(a-\eta)\frac{\partial \Lambda(u)}{\partial u}\big|_{u=a}
 + \Lambda(-a-\eta) \frac{\partial \Lambda(u)}{\partial u}\big|_{u=-a}\big\}\big|_{\{\theta_j=(-1)^ja\}}+E_0\nonumber \\
&=& \phi(2a)\sinh\eta\sum_{j=1}^{2N-1}\big\{\coth(z_j-a-\eta/2)\nonumber\\
 &&+ \coth(z_j+a-\eta/2)\big\}\big|_{\{\theta_j=(-1)^ja\}}+E_0.
\end{eqnarray}

For the system with finite size, we solve the BAEs \eqref{theta2} and obtain the solutions of zero roots. Substituting the
values into \eqref{J1J2redu}, we obtain the eigen-energy of the Hamiltonian (\ref{Ham1}).
The results are given in Table \ref{J1J2redu1}. The eigen-energies can also be obtained by the  exact numerical diagonalization.
We find that the analytical results and numerical ones are consistent with each other very well.
Thus the energy \eqref{J1J2redu} is correct.
\begin{table}[!h]
	\centering
	\caption{ The zero roots and energy spectrum of the system (\ref{Ham1}) with $2N=4$, $a=0.2i$ and $\eta=0.8$.
		Here $E_n$ is the eigen-energy of the $n$-th level and each level is double degenerate. }
	{ \footnotesize
		\begin{tabular}{ccccc}
			\hline
			$z_1$ & $z_2$ & $z_3$ & $E_n$ & $n$ \\ \hline
			$-0.4614i$ & $0$ & $0.4614i$ & $-4.3679$ &  $1$ \\
			$-0.3430i$ & $0.0949i$ & $0.9096i$ & $-3.4531$ &  $2$ \\
			$-0.9096i$ & $-0.0949i$ & $0.3430i$ & $-3.4531$ &  $3$ \\
			$-1.5708i$ & $-0.2291i$ & $0.2291i$ & $-3.2656$ &  $4$ \\
			$-1.0545-1.5708i$ & $0$ & $1.0545-1.5708i$ & $0.6836$ &  $5$ \\
			$-0.8175+0.2545i$ & $-0.2764i$ & $0.8175+0.2545i$ & $3.4531$ &  $6$ \\
			$-0.8175-0.2545i$ & $+0.2764i$ & $0.8175-0.2545i$ & $3.4531$ &  $7$ \\
			$-0.8212$ & $-1.5708i$ & $0.8212$ & $6.9499$ &  $8$ \\
			\hline
	\end{tabular}}\label{J1J2redu1}
\end{table}

\section{The patterns of zero roots}
\setcounter{equation}{0}

Now, we seek the general rules of the solutions of BAEs \eqref{theta2}.
In this paper, we consider the hermitian Hamiltonian, where the model parameter $a$ is pure imaginary and the crossing parameter $\eta$ is real or $\eta\in \mathbb{R}+i\pi$.
We fix the imaginary part of the zero roots in the interval $[-\pi/2, \pi/2)$ because of the periodicity property (\ref{periodic}).
Without losing generality, we set the imaginary parameter $a$ as $a=ib$ and $b$ is real.
From the construction of integrable Hamiltonian (\ref{Ham1}), we know that the inhomogeneous parameters $\{\theta_j\}$ are pure imaginary. In this case, the crossing relation (\ref{crosing-unitarity}) leads to
\begin{eqnarray}
R_{0,j}^{*\dagger_j}(u-\theta_j)=-\sigma_0^yR_{0,j}(-u^*-\eta-\theta_j)\sigma_0^y.
\end{eqnarray}
Substituting the above equation into Eq. (\ref{trans}), we obtain
\begin{eqnarray}
t^\dagger (u)= (-1)^{2N-1}t(-u^*-\eta),
\end{eqnarray}
which gives
\begin{eqnarray}
\Lambda(u)=(-1)^{2N-1}\Lambda^*(-u^*-\eta).\label{conjugate}
\end{eqnarray}
Then we conclude that if the complex number $z_j$ is a root of the BAEs, the $-z_j^*$ must be another root.
Thus the zero roots form the pairing solutions which have the same imaginary part but the real parts are opposite, i.e.,
\begin{eqnarray}\label{zroot}
\mathrm{Re}(z_j)+\mathrm{Re}(z_l)=0,\quad \mathrm{Im}(z_j)=\mathrm{Im}(z_l).
\end{eqnarray}
The zero roots are distributed symmetrically about the imaginary axis.

The more detailed distribution of zero roots could be obtained with the help of Bethe roots.
The functional identity (\ref{theta1}) allows us to parameterize the eigenvalue $\Lambda(u)$ as the inhomogeneous $T-Q$ relation \cite{Book}
\begin{eqnarray}\label{TQ}
\Lambda(u)Q(u)=e^ua(u)Q(u-\eta)-e^{-u-\eta}d(u)Q(u+\eta)-c(u)a(u)d(u),
\end{eqnarray}
where the $Q(u)$ and $c(u)$ are given by
\begin{eqnarray}
&&Q(u)=\prod_{j=1}^{2N}\frac{\sinh(u-\lambda_j)}{\sinh\eta},  \nonumber\\
&&c(u)=e^{u-2N\eta+\sum_{l=1}^{2N}(\theta_l-\lambda_l)}-e^{-u-\eta-\sum_{l=1}^{2N}(\theta_l-\lambda_l)},
\end{eqnarray}
and $\{\lambda_j\}$ are the Bethe roots, which should satisfy the BAEs
\begin{eqnarray}\label{orBAEs-inh}
&&e^{\lambda_j}a(\lambda_j)Q(\lambda_j-\eta)-e^{-\lambda_j-\eta}d(\lambda_j)Q(\lambda_j+\eta)\nonumber\\
&&-c(\lambda_j)a(\lambda_j)d(\lambda_j)=0, \quad j=1,\cdots,2N.
\end{eqnarray}
Putting $\lambda_j=i u_j-\eta/2$ and taking the stagger limit $\theta_j=(-1)^ja$, we rewrite the above BAEs as
\begin{eqnarray}\label{BAEs2}
&&e^{iu_j}\frac{\sin^N(u_j+b-\frac{1}{2}i\eta) \sin^N(u_j-b-\frac{1}{2}i\eta)}{\sin^N(u_j+b+\frac{1}{2}i\eta) \sin^N(u_j-b+\frac{1}{2}i\eta)}\nonumber\\
&&=e^{-iu_j}\prod_{l=1}^{2N}\frac{\sin(u_j-u_l-i\eta)}{\sin(u_j-u_l+i\eta)}+2i\,e^{-N\eta}\sin\big(u_j-\sum_{l=1}^{2N}{u_l}\big) \nonumber \\
&&\times\frac{\sin^N(u_j+b-\frac{1}{2}i\eta) \sin^N(u_j-b-\frac{1}{2}i\eta)}{\prod_{l=1}^{2N}\sin(u_j-u_l+i\eta)},\quad j=1,\cdots,2N.
\end{eqnarray}
When $\eta$ is positive real, for a complex $u_j$ with a negative imaginary part, we have
\begin{eqnarray}
\bigg|\sin(u_j\pm b-\frac{1}{2}i\eta)\bigg|\textgreater\bigg|\sin(u_j\pm b+\frac{1}{2}i\eta)\bigg|.
\end{eqnarray}
This indicates that the module of the left hand side of BAEs (\ref{BAEs2}) tends to infinity exponentially when $N\to\infty$. To keep the equality, the denominator of the right-hand side of BAEs (\ref{BAEs2}) must tend to zero in this limit, which gives that $u_j-u_l+i\eta\to 0$.
From the $T-Q$ relation (\ref{TQ}), we know that the zero roots $\{z_j-\frac\eta2\}$ and $\{iu_j-\frac\eta2\}$ of the term $\Lambda(u)Q(u)$  are undistinguishable, so $\{u_j\}$ are symmetric about the real axis since $\{z_j\}$ are symmetric about the imaginary axis from (\ref{zroot}). Therefore the general complex solutions of the Bethe roots form strings
\begin{eqnarray}
u_j=u_{j0}+i \eta\big(\frac{n+1}{2}-j\big)+o(e^{-\delta N}), \quad j=1,\cdots,n,
\end{eqnarray}
where $u_{j0}$ indicates the position of the $n$-string in the real axis and $o(e^{-\delta N})$ stands for a small finite size correction.

Now we can determine the pattern of zero roots $\{z_j\}$. Putting $z_j=ix_j$ and taking the zero root $ix_j-\frac\eta2$ into Eq.(\ref{TQ}), we obtain
\begin{eqnarray}\label{BAEs3}
&&e^{ix_j}\frac{\sin^N(x_j+b-\frac{1}{2}i\eta) \sin^N(x_j-b-\frac{1}{2}i\eta)}{\sin^N(x_j+b+\frac{1}{2}i\eta) \sin^N(x_j-b+\frac{1}{2}i\eta)}\nonumber\\
&&=e^{-ix_j}\prod_{l=1}^{2N}\frac{\sin(x_j-u_l-i\eta)}{\sin(x_j-u_l+i\eta)}+ 2i\,e^{-N\eta}\sin\big(x_j-\sum_{l=1}^{2N}{u_l}\big)\nonumber \\
&&\times\frac{\sin^N(x_j+b-\frac{1}{2}i\eta) \sin^N(x_j-b-\frac{1}{2}i\eta)}{\prod_{l=1}^{2N}\sin(x_j-u_l+i\eta)},
\quad j=1,\cdots,2N.
\end{eqnarray}
A similar discussion can then proceed. For the $x_j$ with a negative imaginary part, the equation (\ref{BAEs3}) leads to the relation between the zero root and  Bethe root as $x_j-u_l+i\eta\to 0$ when $N$ tends to infinity. One should note that the two sets of roots could not be equal and the zero roots are lower in the complex plane than the Bethe roots. Combined with the fact that the $\{x_j\}$ are symmetric about the real axis from (\ref{zroot}), we arrive at the similar statement that, for $x_i$ with a positive imaginary part, $x_i-u_k-i\eta\to 0$ with $N\to\infty$ and the corresponding zero roots are higher in the complex plane than the Bethe roots. Thus the above analysis determines the pattern of zero roots $\{x_j\}$ as
\begin{eqnarray}
\mathrm{Im}(x_j)=\pm \frac{1+n}2 \eta+o(e^{-\delta N}), \quad n=1,2,\cdots.
\end{eqnarray}
Substituting them into $z_j=ix_j$, we obtain
\begin{eqnarray}\label{zroot1}
\mathrm{Re}(z_j)=\pm \frac{1+n}2 \eta+o(e^{-\delta N}), \quad n=1,2,\cdots.
\end{eqnarray}
The above conclusion also holds for $\eta\in \mathbb{R}+i\pi$ by replacing $\eta$ with $\mathrm{Re}(\eta)$.

\section{The nearly degenerate states}

\begin{figure}[t]
\begin{center}
\includegraphics[width=5cm]{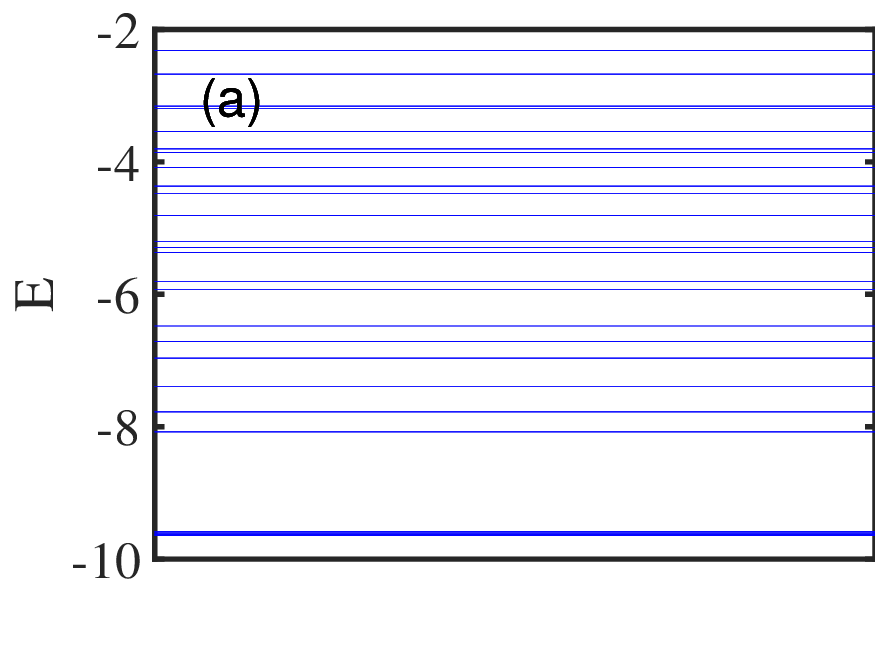}
\includegraphics[width=5cm]{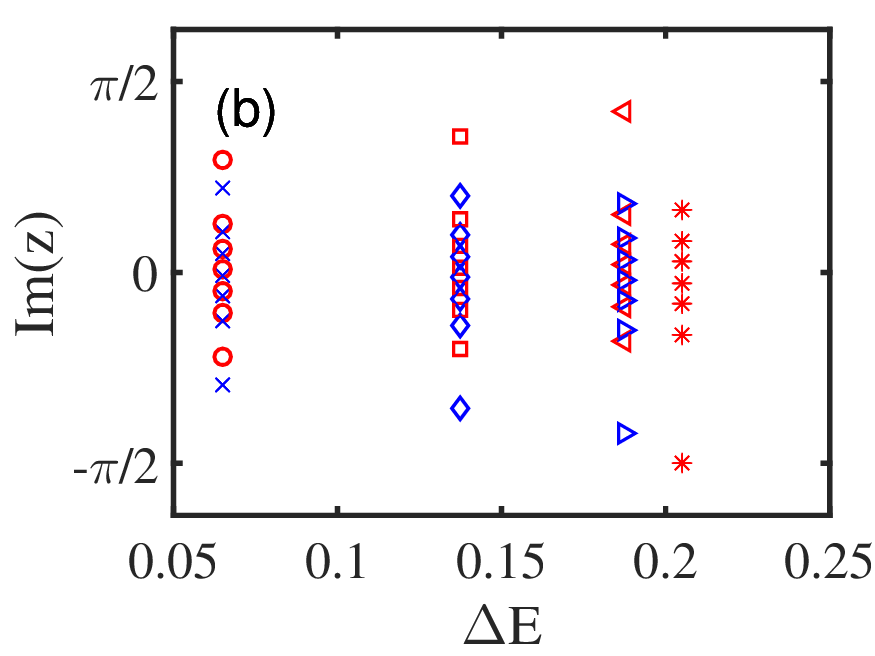}
\includegraphics[width=5cm]{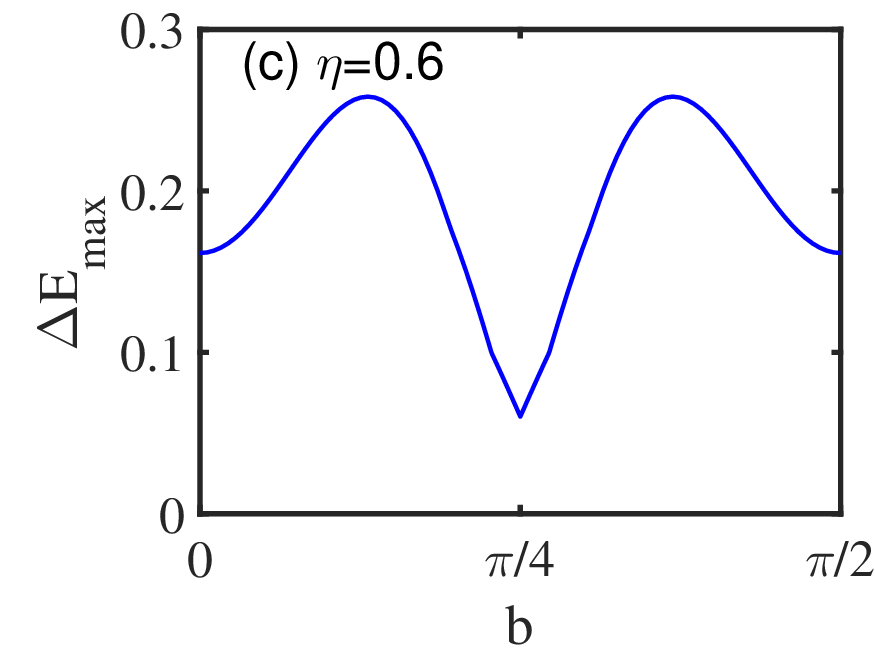}
\caption{(a) The system's energy levels with $2N=8$, $b=0.2$ and $\eta=0.8$, where we have omitted some levels at high excited states.
(b) The distribution of zero root at the nearly degenerate states for different energy difference $\Delta E$ with $2N=8$, $b=0.2$ and $\eta=0.6$.
There are 14 nearly degenerate states and 7 energy differences. The set of solutions of zero roots are denoted by different colors.
For example, considering the double degeneracy, there are $(4N-2)/2=7$ sets of different zero roots for the nearly degenerate states at $2N=8$ case. We plot the 7 sets of zero roots with different colors in Fig.\ref{fig-deE1}(b) for a clearer explanation.
(c) The energy difference between the highest nearly degenerate state and the ground state $\Delta E_{max}$ versus the model parameter $a$ with $2N=8$ and $\eta=0.6$.}\label{fig-deE1}
\end{center}
\end{figure}

By carefully analyzing the energy spectrum, we find an interesting phenomenon: some nearly degenerate states exist in the regime of $\eta\in \mathbb{R}$, where the NN couplings among the $z$-direction are ferromagnetic.
The energy spectrum is shown in Fig.\ref{fig-deE1}(a). One sees that the energy levels can be divided into two parts. There is a big gap between the lower and upper energy levels. Further analysis gives that the patterns of zero points in these two regimes are different.
In the lower regime, all the zero roots are pure imaginary and are asymmetric around the origin. The related states are the nearly degenerate states.

From the numerical results of the system with finite size, we find that there are $4N$ sets of zero roots lying on the imaginary axis, in which 2 sets correspond to the ground states. The other $4N-2$ sets correspond to the nearly degenerate states. The degeneracy of the ground state is $2$.
In Fig.\ref{fig-deE1}(b), we show the patterns of zero roots in the nearly degenerate states.
Because all the zero roots are located on the imaginary axis, it is not necessary to show them in the complex plane.
Thus we choose the lateral axis of Fig.\ref{fig-deE1}(b) as the energy difference $\Delta E=E_d-E_{1g}$ instead of the real axis, where
$E_d$ is the energy of nearly degenerate state and $E_{1g}$ is the ground state energy.

To further investigate the physical properties of the nearly degenerate states, we calculate the spin texture of these states by the exact numerical diagonalization.
Table \ref{dt} shows that the ground states and the nearly degenerate states can be regarded as the superpositions of domain walls or kinks,
which are generated by continuously flipping some spins from the all spin-up state (or the all spin-down state),
while the high excited states can not be. The low-lying states, i.e., the ground and nearly degenerate states, have two domain walls. One is fixed between sites $2N$ and $1$ due to the antiperiodic boundary.
The other can be located between sites $j$ and $j+1$. This in total gives $2\times2N$ configurations.
Subtracting two degenerated ground states, we have $4N-2$ nearly degenerate states, which is consistent with the
numerical results.

\begin{table}[!h]
\centering
\caption{The projections $\alpha_{ij}=\langle F_j|\psi_i\rangle$ and the error $\delta\equiv \big||\psi_i\rangle-\sum_{j=1}^{8}\alpha_{ij}|F_j\rangle\big|$
with $2N=4$, $a=0.2i$ and $\eta=2$. Here, $\{|\psi_i\rangle, i=1,\cdots,16\}$ are the eigenstates. Among them, the
first two are the ground states, from third to 8-th are the nearly degenerate states, and the rest ones are the high excited states.
$\{|F_j\rangle=\prod_{k=1}^{j-1}\sigma^x_{(k-1) mod (4)+1}|\uparrow\uparrow\uparrow\uparrow\rangle,j=1,\cdots,8\}$ are the approximate basis vectors of 8-dimensional low-lying states.
The values of $\delta$ at the low-lying states are much smaller than those at the high excited states. }\label{dt}
{ \footnotesize
\begin{tabular}{l|llll}
\hline
   $\alpha_{ij}$   & $j=1$   & $j=2$  & $j=3$  & $j=4$  \\ \hline
$i=1$  & $0.0087-0.0000i$ & $0.4718-0.1416i$ & $0.0087-0.0000i$ & $0.4718-0.1416i$ \\
$i=2$  & $0.4926+0.0000i$ & $-0.0083+0.0025i$ & $0.4926+0.0000i$ & $-0.0083+0.0025i$ \\
$i=3$  & $-0.0000-0.0000i$ & $0.1609+0.5121i$ & $-0.0041+0.2356i$ & $0.1343-0.3603i$ \\
$i=4$  & $-0.0000+0.0000i$ & $0.0742-0.4350i$ & $-0.1622+0.1233i$ & $0.3358-0.3775i$ \\
$i=5$  & $-0.0003-0.0000i$ & $0.0865-0.0343i$ & $-0.4582+0.4296i$ & $-0.2278+0.1911i$ \\
$i=6$  & $-0.7011-0.0000i$ & $-0.0000+0.0000i$ & $0.0002-0.0002i$ & $0.0001-0.0001i$ \\
$i=7$  & $0.0030+0.0000i$ & $-0.4606+0.1871i$ & $-0.0030-0.0000i$ & $0.4606-0.1871i$ \\
$i=8$  & $0.4972+0.0000i$ & $0.0028-0.0011i$ & $-0.4972-0.0000i$ & $-0.0028+0.0011i$ \\
$i=9$  & $-0.0045+0.0000i$ & $-0.0826+0.0198i$ & $-0.0045+0.0000i$ & $-0.0826+0.0198i$ \\
$i=10$  & $0.0850+0.0000i$ & $-0.0044+0.0011i$ & $0.0850-0.0000i$ & $-0.0044+0.0011i$ \\
$i=11$  & $-0.0038-0.0000i$ & $0.0210+0.0302i$ & $0.0431-0.0244i$ & $-0.0660-0.0173i$ \\
$i=12$  & $0.0900+0.0000i$ & $0.0064+0.0065i$ & $0.0139+0.0033i$ & $0.0056+0.0079i$ \\
$i=13$  & $0.0191-0.0000i$ & $-0.0258-0.0244i$ & $-0.0568-0.0203i$ & $-0.0395-0.0406i$ \\
$i=14$  & $-0.0000-0.0000i$ & $0.0715+0.0259i$ & $-0.0400+0.0240i$ & $0.0054-0.0222i$ \\
$i=15$  & $0.0529+0.0000i$ & $0.0043+0.0003i$ & $-0.0529-0.0000i$ & $-0.0043-0.0003i$ \\
$i=16$  & $-0.0043-0.0000i$ & $0.0528+0.0034i$ & $0.0043-0.0000i$ & $-0.0528-0.0034i$ \\
\hline
\end{tabular}
\begin{tabular}{llll|l}
\hline
 $j=5$ &$j=6$  &  $j=7$&  $j=8$ & $\delta$\\ \hline
$0.0087$ & $0.4718-0.1416i$ & $0.0087+0.0000i$ & $0.4718-0.1416i$ &  $0.1702$ \\
 $0.4926$ & $-0.0083+0.0025i$ & $0.4926-0.0000i$ & $-0.0083+0.0025i$ &  $0.1702$ \\
 $0.0000$ & $-0.1609-0.5121i$ & $0.0041-0.2356i$ & $-0.1343+0.3603i$ &  $0.1302$ \\
 $0.0000$ & $-0.0742+0.4350i$ & $0.1622-0.1233i$ & $-0.3358+0.3775i$ &  $0.1302$ \\
 $0.0003$ & $-0.0865+0.0343i$ & $0.4582-0.4296i$ & $0.2278-0.1911i$ &  $0.1302$ \\
 $0.7011$ & $0.0000-0.0000i$ & $-0.0002+0.0002i$ & $-0.0001+0.0001i$ &  $0.1302$ \\
 $0.0030$ & $-0.4606+0.1871i$ & $-0.0030+0.0000i$ & $0.4606-0.1871i$ &  $0.1062$ \\
 $0.4972$ & $0.0028-0.0011i$ & $-0.4972+0.0000i$ & $-0.0028+0.0011i$ &  $0.1062$ \\
 $-0.0045$ & $-0.0826+0.0198i$ & $-0.0045-0.0000i$ & $-0.0826+0.0198i$ &  $0.9854$ \\
 $0.0850$ & $-0.0044+0.0011i$ & $0.0850+0.0000i$ & $-0.0044+0.0011i$ &  $0.9854$ \\
 $0.0038$ & $-0.0210-0.0302i$ & $-0.0431+0.0244i$ & $0.0660+0.0173i$ &  $0.9915$ \\
 $-0.0900$ & $-0.0064-0.0065i$ & $-0.0139-0.0033i$ & $-0.0056-0.0079i$ &  $0.9915$ \\
 $-0.0191$ & $0.0258+0.0244i$ & $0.0568+0.0203i$ & $0.0395+0.0406i$ &  $0.9915$ \\
 $0.0000$ & $-0.0715-0.0259i$ & $0.0400-0.0240i$ & $-0.0054+0.0222i$ &  $0.9915$ \\
 $0.0529$ & $0.0043+0.0003i$ & $-0.0529-0.0000i$ & $-0.0043-0.0003i$ &  $0.9943$ \\
 $-0.0043$ & $0.0528+0.0034i$ & $0.0043-0.0000i$ & $-0.0528-0.0034i$ &  $0.9943$ \\
\hline
\end{tabular}}
\end{table}

Now, we consider the relation between the nearly degenerate states and the interactions.
Define $\Delta E_{max}=max(E_d)-E_{1g}$, where $max(E_d)$ is the maximal energy of the nearly degenerate states
and $E_{1g}$ is the ground state energy. The energy difference $\Delta E_{max}$ versus the model parameter $a=ib$ is shown in Fig.\ref{fig-deE1}(c).
From it, we see that the $\Delta E_{max}$ changes with the changing of NN, NNN and chiral three-spin interactions and reaches
its minimum at the point of $a=i\pi/4$.

Last, we shall note that the gaps among the nearly degenerate states tend to zero with the increasing of system size.
In the thermodynamic limit, these nearly degenerate states become the ground state.

\section{Thermodynamic limit with $\eta\in \mathbb{R}$}
\setcounter{equation}{0}

Since we have known the zero roots distribution of the BAEs, it is now possible to calculate the physical quantities in the thermodynamic limit. Based on the $t-\theta$ scheme proposed in \cite{Qiao21,Le21}, we choose the inhomogeneity parameters $\{\theta_j\}$ as auxiliary ones to calculate the physical quantities such as the ground state energy and the elementary excitations of the system.
We first consider the regime of $\eta$ is real.
From the previous derivation, we know that $\{\theta_j\}$ are imaginary because that $a$ is imaginary.

\subsection{The ground state}

\begin{figure}[t]
\begin{center}
\includegraphics[height=5.5cm,width=7cm]{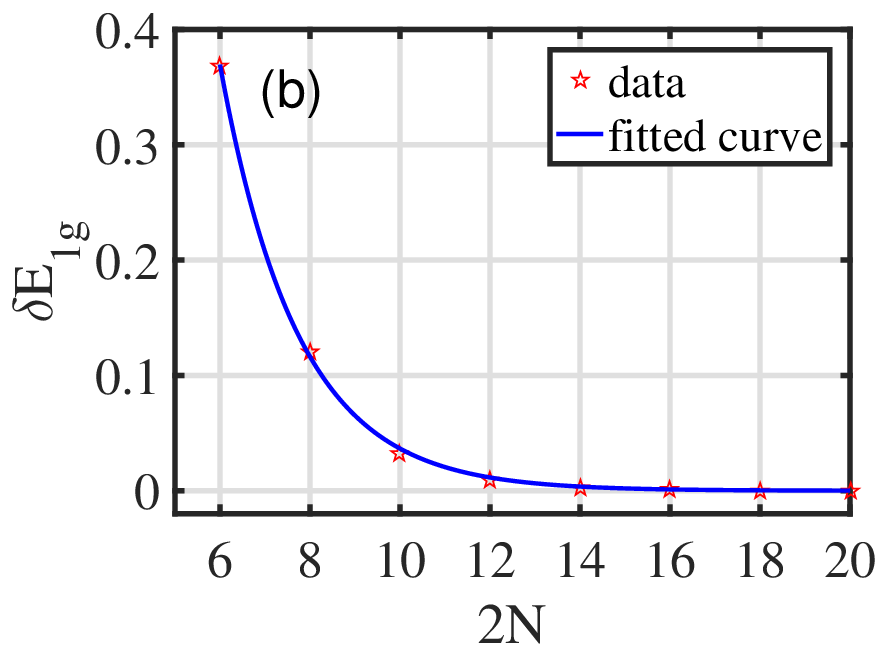}
\caption{ The finite size scaling behavior of the quantity $\delta E_{1g}=E_{1g}-E_{1gd}$, where
$E_{1g}$ is that obtained from the analytic result (\ref{grounden1}) and
$E_{1gd}$ is the ground state energy calculated by the exact numerical diagonalization with finite system size. The data of $\delta E_{1g}$ can be fitted as $\delta E_{1g}=11.86e^{-0.5782\times2N}$ with $b=0.2$ and $\eta=0.6$,
which is exponentially decreasing and tends to zero in the thermodynamic limit.
}\label{fig-E1}
\end{center}
\end{figure}

At the ground state, all roots $\{z_j\}$ take imaginary values for the imaginary $\{\theta_j\}$. It is convenient to put $\theta_j=i\phi_j$ and $z_j=ix_j$, where $\phi_j$ and $x_j$ take real values. Taking the logarithm and considering the thermodynamic limit $N\to\infty$ of Eq.(\ref{theta2}), we obtain
\begin{eqnarray}\label{theta3}
&&\ln|\Lambda_0^2|+2N\int\beta_1(\phi-x)\rho(x)dx=\ln|\sinh^{-4N}\eta|+2N\int\beta_2(\phi-x)\sigma(x)dx,
\end{eqnarray}
where $\beta_n(x)=\ln[\sin(x-in\eta/2)\sin(x+in\eta/2)]$, $\rho(x)$ and $\sigma(x)$ are the density of $\{x_j\}$  and $\{\phi_j\}$, respectively.
Taking the derivative of Eq.(\ref{theta3}) with respect to $\phi$, we have
\begin{eqnarray}\label{theta4}
\int_{-\frac\pi2}^{\frac\pi2} b_1(\phi-x)\rho(x)dx=\int_{-\frac\pi2}^{\frac\pi2} b_2(\phi-x)\sigma(x)dx,
\end{eqnarray}
in which $b_n(x)=2\sin(2x)/[(\cos n\eta-\cos2x)]$. Introduce the Fourier transformation
\begin{eqnarray}
f(x)=\frac{1}\pi\sum_{\omega=-\infty}^{+\infty}\tilde{f}(\omega)e^{i2\omega x}, \quad
\tilde{f}(\omega)=\int_{-\frac\pi2}^{\frac\pi2}f(x)e^{-i2\omega x} dx.
\end{eqnarray}
The Fourier transformation of Eq.(\ref{theta4}) reads
\begin{eqnarray}
\tilde{b}_1(\omega)\tilde{\rho}(\omega)=\tilde{b}_2(\omega)\tilde{\sigma}(\omega),
\end{eqnarray}
where $\tilde{b}_n(\omega)=-i2\pi sign(\omega)e^{-n\eta\lvert\omega\rvert}$. Because the total number for zero roots $\{z_j\}$ is $2N-1$, thus the normalization of
zero roots density $\rho(x)$ should satisfy $\int_{-\frac\pi2}^{\frac\pi2}\rho(x)dx=\frac{2N-1}{2N}$.
In the thermodynamic limit, the density of inhomogeneous parameters $\{\theta_j=(-1)^ja, j=1,\cdots, 2N\}$ becomes $\sigma(x)=\frac{1}{2}[\delta(x-b)+\delta(x+b)]$.
Taking the Fourier transformation of $\sigma(x)$, we obtain $\tilde{\sigma}(\omega)=\frac{1}{2}(e^{i2\omega b}+e^{-i2\omega b})$.
Therefore, the solution of zero roots density is
\begin{eqnarray}
\tilde{\rho}(\omega)=\left\{
\begin{aligned}
&e^{-\eta\lvert\omega\rvert}\cos(2\omega b),\quad \omega=\pm1,\pm2,\cdots,\pm\infty, &\\
&1-\frac1{2N},\quad\omega=0.&
\end{aligned}
\right.
\end{eqnarray}
Taking the inverse Fourier transformation, we obtain
\begin{eqnarray}\label{groundrho1}
  \rho(x)&=&\frac{1}{\pi} \bigg \{ \frac{1-e^{-\eta}\cos(2x+2b)}{1-2e^{-\eta}\cos(2x+2b)+e^{-2\eta}}
  +\frac{1-e^{-\eta}\cos(2x-2b)}{1-2e^{-\eta}\cos(2x-2b)+e^{-2\eta}} \bigg \}\nonumber\\
  &&-\frac1\pi-\frac1{2N\pi}.
\end{eqnarray}
The ground state energy can be calculated as
\begin{eqnarray}\label{grounden1}
E_{1g}&=&2N\phi(2a)\sinh\eta \int_{-\frac\pi2}^{\frac\pi2} [\coth(ix-ib-\eta/2) + \coth(ix+ib-\eta/2)]\rho(x)dx+E_0 \nonumber\\
 &=&-2N\cosh\eta-N\frac{\cosh\eta \sin^2(2b)}{\sinh^2\eta}+\frac{\cosh(2\eta)-\cos(4b)}{\sinh\eta}.
\end{eqnarray}
Now, we check the correctness of the result \eqref{grounden1}. For the system (\ref{Ham1}) with finite size, we obtain the ground state energy $E_{1gd}$ by using the exact diagonalization method.
Define $\delta E_{1g}=E_{1g}-E_{1gd}$, where $E_{1g}$ is the ground state energy calculated from the analytic expression (\ref{grounden1}).
We note that both the values of $E_{1gd}$ and $\delta E_{1g}$ are dependent on the system size.
Then we take the finite size scaling analysis and the results are shown in
Fig.\ref{fig-E1}. We find that the data of $\delta E_{1g}$ can be fitted as $\delta E_{1g}=11.86e^{-0.5782\times2N}$, where $b=0.2$ and $\eta=0.6$.
Thus $\delta E_{1g}$ is exponentially decreasing with the increasing of system size.
$\delta E_{1g}\rightarrow 0$ if $N \rightarrow \infty$. Therefore, the analytic expression (\ref{grounden1})
gives the ground state energy in the thermodynamic limit.

\subsection{Elementary excitation}

\begin{figure}[t]
\begin{center}
\includegraphics[width=5cm]{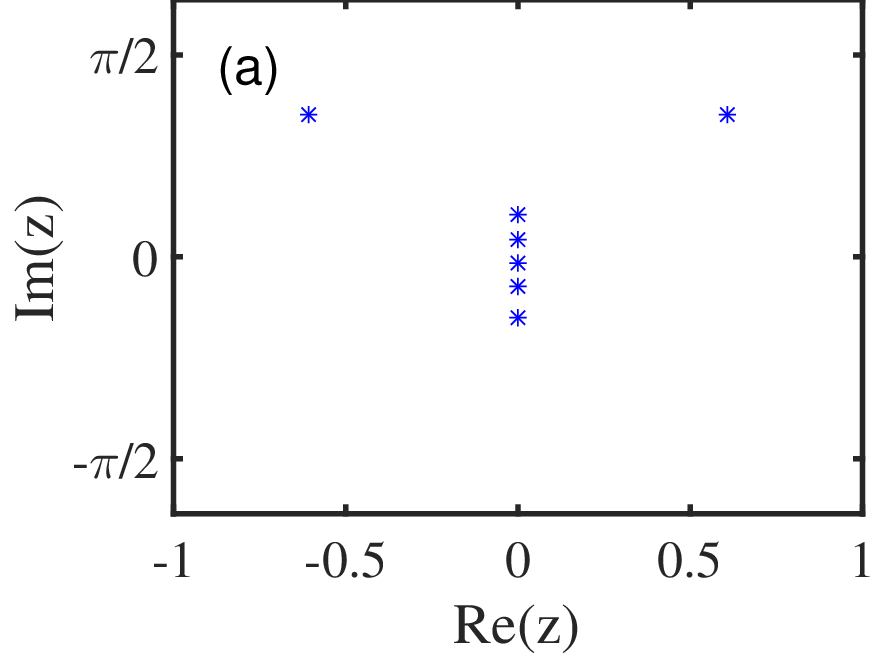}
\includegraphics[height=4.2cm,width=5cm]{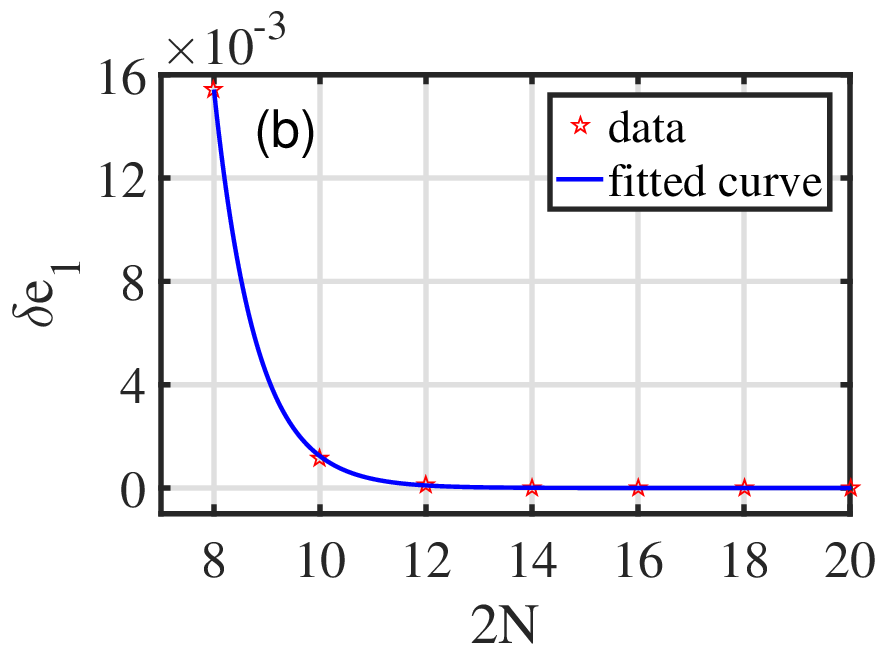}
\includegraphics[width=5cm]{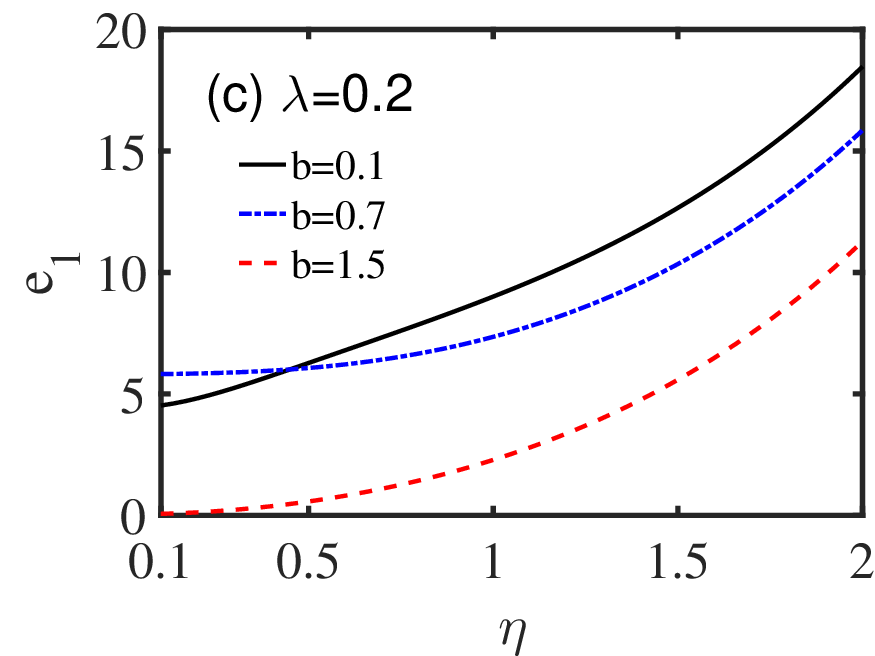}
\caption{(a) The distribution of zero roots at the low-lying excited state with $2N=8$, $b=0.2$ and $\eta=0.6$.
(b) The finite size scaling behavior of $\delta e_{1}=e_{1}-e_{1d}$, where $e_1$ is the excited energy obtained by the expression (\ref{e1})
and $e_{1d}$ is that computed by using the exact numerical diagonalization with finite system size. Here $b=0.75$ and $\eta=1$.
The data can be fitted as $\delta e_{1}=376.8e^{-1.263\times2N}$, which tends to zero when $N\rightarrow \infty$.
(c) The excited energies with given values of $b$ versus the anisotropic parameter $\eta$, where $n=2$ and $\lambda=0.2$.
}\label{fig-ee1}
\end{center}
\end{figure}

Now we study the elementary excitation. From the general constraints of zero roots (\ref{zroot}), the distribution of $\{z_j\}$ for the simplest excited state
can be described by $2N-3$ imaginary roots plus one conjugate pair. The extra conjugate paired are
\begin{eqnarray}\label{eez1}
z_{2N-2}=i\lambda+\frac{n\eta}2+o(e^{-\delta N}),\nonumber\\
z_{2N-1}=i\lambda-\frac{n\eta}2+o(e^{-\delta N}),
\end{eqnarray}
where $\lambda$ is real and $n\geq2$. The distribution of zero roots for such an excitation with $2N=8$ is shown in Fig.\ref{fig-ee1}(a).
Substituting all the zero roots into BAEs (\ref{theta2}) and considering the thermodynamic limit, we obtain
\begin{eqnarray}
&&\ln\Lambda_0^2+2N\int\beta_1(\phi-x)\rho(x)dx+\beta_{n+1}(\phi-\lambda)+\beta_{n-1}(\phi-\lambda)\nonumber\\
&&=\ln(-\sinh^{-4N}\eta)+2N\int\beta_2(\phi-x)\sigma(x)dx.
\end{eqnarray}
The derivative with respect to $\phi$ gives
\begin{eqnarray}\label{eeb}
&&2N\int_{-\frac\pi2}^{\frac\pi2} b_1(\phi-x)\rho_1(x)dx+b_{n+1}(\phi-\lambda)+b_{n-1}(\phi-\lambda)\nonumber\\
&&=2N\int_{-\frac\pi2}^{\frac\pi2} b_2(\phi-x)\sigma(x)dx.
\end{eqnarray}
Taking the Fourier transformation of (\ref{eeb}), we obtain
\begin{eqnarray}\label{e222eb}
2N \tilde{b}_1(\omega)\tilde{\rho}_1(\omega)+e^{-i2\omega\lambda}\tilde{b}_{n+1}(\omega)+e^{-i2\omega\lambda}\tilde{b}_{n-1}(\omega)=2N \tilde{b}_2(\omega)\tilde{\sigma}(\omega).
\end{eqnarray}
With the help of normalization $\int_{-\frac\pi2}^{\frac\pi2}\rho(x)dx=\frac{2N-3}{2N}$ and $\sigma(x)=\frac{1}{2}[\delta(x-b)+\delta(x+b)]$,
we obtain the density of zero roots $\tilde{\rho}_1(\omega)$ as
\begin{eqnarray}
\tilde{\rho}_1(\omega)=\left\{
\begin{aligned}
&e^{-\eta\lvert\omega\rvert}\cos(2\omega b)-\frac{e^{-i2\omega\lambda}}{2N}(e^{-n\eta\lvert\omega\rvert}+e^{-(n-2)\eta\lvert\omega\rvert}),
\quad \omega=\pm1,\pm2,\cdots,\pm\infty,& \\
&1-\frac3{2N},\quad\omega=0.&
\end{aligned}
\right.
\end{eqnarray}
The inverse Fourier transformation of $\tilde{\rho}_1(\omega)$ gives
\begin{eqnarray}
  \rho_1(x)&=&-\frac{1}{N\pi} \bigg \{ \frac{1-e^{-n\eta}\cos(2x-2\lambda)}{1-2e^{-n\eta}\cos(2x-2\lambda)+e^{-2n\eta}}
  +\frac{1-e^{-(n-2)\eta}\cos(2x-2\lambda)}{1-2e^{-(n-2)\eta}\cos(2x-2\lambda)+e^{-2(n-2)\eta}} \bigg \}\nonumber\\
  &&+\frac1{N\pi}+\rho(x),
\end{eqnarray}
where $\rho(x)$ is given by Eq.(\ref{groundrho1}).
Substituting the density of zero roots into Eq.(\ref{J1J2redu}), we obtain elementary excitation energy
\begin{eqnarray}\label{e1}
  e_1(\lambda) &=&2N\phi(2a)\sinh\eta \int_{-\frac\pi2}^{\frac\pi2} [\coth(ix-ib-\eta/2) + \coth(ix+ib-\eta/2)][\rho_1(x)-\rho(x)]dx \nonumber\\
  &&+\phi(2a)\sinh\eta[\coth(\frac{n-1}2\eta+i\lambda-ib)+\coth(\frac{n-1}2\eta+i\lambda+ib)\nonumber\\
  &&+\coth(-\frac{n+1}2\eta+i\lambda-ib)+\coth(-\frac{n+1}2\eta+i\lambda+ib)]\nonumber\\
  &=&\frac{\cosh(2\eta)-\cos(4b)}{\sinh\eta} \bigg[\frac{\sinh(n-1)\eta}
  {\cosh(n-1)\eta-\cos(2\lambda+2b)}\nonumber\\
  &&+\frac{\sinh(n-1)\eta}{\cosh(n-1)\eta-\cos(2\lambda-2b)}\bigg].
\end{eqnarray}

Now, we check the correctness of Eq.(\ref{e1}).
Define $\delta e_{1}=e_{1}-e_{1d}$, where $e_1$ is the excited energy obtained by the expression (\ref{e1})
and $e_{1d}$ is that computed by using the exact numerical diagonalization with finite system size.
The finite size scaling behavior of $\delta e_{1}$ is shown in Fig.\ref{fig-ee1}(b).
We see that the data can be fitted as $\delta e_{1}=376.8e^{-1.263\times2N}$.
In the thermodynamic limit, $\delta e_{1}$ tends to zero. Thus the analytic result (\ref{e1}) is correct.
The excited energies with given values of $b$ versus the anisotropic parameter $\eta$ are plotted in Fig.\ref{fig-ee1}(c).
We see that the excited energy increases with the increasing of $\eta$.

Substituting the distribution of zero roots (\ref{eez1}) at this kind of excited state into Eq.(\ref{kz}), we obtain the momentum carried by the elementary excitation as
\begin{eqnarray}\label{k1}
k_1(\lambda) &=& -i2N\int_{-\frac\pi2}^{\frac\pi2}\ln\frac{\sinh(a+ix-\frac\eta2)}{\sinh(a-ix-\frac\eta2)}[\rho_1(x)-\rho(x)]dx\nonumber\\
    &&-i\ln\frac{\sinh(a+i\lambda+\frac{n\eta}2-\frac\eta2)\sinh(a+i\lambda-\frac{n\eta}2-\frac\eta2)}
    {\sinh(a-i\lambda-\frac{n\eta}2-\frac\eta2)\sinh(a-i\lambda+\frac{n\eta}2-\frac\eta2)} {~~}mod\,(2\pi)\nonumber\\
    &=&4\sum_{\omega=1}^{\infty}\frac{\sin(2\omega\lambda)}\omega\cos(2\omega b)e^{-n\eta\omega}\cosh(\eta\omega)\nonumber\\
    &&+\frac{i}2\bigg[ \ln\frac{\sin(b+\lambda+i\frac{n-1}2\eta)\sin(b+\lambda-i\frac{n+1}2\eta)}
    {\sin(b-\lambda+i\frac{n-1}2\eta)\sin(b-\lambda-i\frac{n+1}2\eta)} \nonumber\\
    &&-\ln\frac{\sin(b+\lambda-i\frac{n-1}2\eta)\sin(b+\lambda+i\frac{n+1}2\eta)}
    {\sin(b-\lambda-i\frac{n-1}2\eta)\sin(b-\lambda+i\frac{n+1}2\eta)}\bigg] {~~}mod\,(2\pi).
\end{eqnarray}

\section{Thermodynamic limit with $\eta\in \mathbb{R}+i\pi$}
\setcounter{equation}{0}

In this section, we study the physical quantities in the regime of $\eta\in \mathbb{R}+i\pi$.
The patterns of zero roots of BAEs are given by Eq.(\ref{zroot1}) with the replacing $\eta$ by $\mathrm{Re}(\eta)$.
For simplicity, we define $\eta_+=\eta-i\pi$, then the patterns read
\begin{eqnarray}
\mathrm{Re}(z_j)=\pm \frac{1+n}2 \eta_++o(e^{-\delta N}), \quad n=1,2,\cdots.
\end{eqnarray}
Without losing generality, we suppose $b\in(0,\pi/2)$.

\subsection{Ground state and quantum phase transition}

At the ground state, the pattern of zero toots $\{z_j\}$  includes $N-1$ conjugate pairs and one pure imaginary solution, i.e.,
\begin{eqnarray}\label{zroot2}
&&z_{2j-1}=ix_{2j-1}+\eta_++o(e^{-\delta N}),\quad z_{2j}=ix_{2j}-\eta_++o(e^{-\delta N}),\quad j=1,\cdots,N-1,\nonumber\\
&&z_{2N-1}=i\mu,
\end{eqnarray}
where $\{x_{j}\}$ and $\mu$ are real. Substituting the pattern (\ref{zroot2}) into Eq.(\ref{theta2}), taking the logarithm and considering the thermodynamic limit, we obtain
\begin{eqnarray}
&&\ln|\Lambda_0^2|+\gamma_1(\phi-\mu)+2N\int[\gamma_1(\phi-x)+\gamma_3(\phi-x)]\rho_2(x)dx\nonumber\\
&&=\ln|\sinh^{-4N}\eta_+|+2N\int\beta_2(\phi-x)\sigma(x)dz,
\end{eqnarray}
in which $\gamma_n(x)=\ln[\cos(x-in\eta_+/2)\cos(x+in\eta_+/2)$], and $\rho_2(x)$ and $\sigma(x)$ denote the density of $\{x_j\}$  and $\{\phi_j\}$, respectively. Taking the derivative, we have
\begin{eqnarray}\label{zrwwoot2}
-c_1(\phi-\mu)-2N\int_{-\frac\pi2}^{\frac\pi2} [c_1(\phi-x)+c_3(\phi-x)]\rho_2(x)dx =2N\int_{-\frac\pi2}^{\frac\pi2} b_2(\phi-x)\sigma(x)dx,
\end{eqnarray}
where $c_n(x)=\tan(x+in\eta_+/2)+\tan(x-in\eta_+/2)$. The Fourier transformation of Eq.\eqref{zrwwoot2} gives
\begin{eqnarray}
-e^{-i2\omega\mu}\tilde{c}_1(\omega)-2N[\tilde{c}_1(\omega)+\tilde{c}_3(\omega)]\tilde{\rho}_2(\omega) =2N\tilde{b}_2(\omega)\tilde{\sigma}(\omega),
\end{eqnarray}
where $\tilde{c}_n(\omega)=(-1)^{\omega}sign(\omega)2\pi ie^{-n\eta\lvert\omega\rvert}$.
With the help of normalization $\int_{-\frac\pi2}^{\frac\pi2}\rho(x)dx=\frac12-\frac1{2N}$, we obtain the solution of zero roots density
\begin{eqnarray}
\tilde{\rho}_2(\omega)=\left\{
\begin{aligned}
&-\frac{\frac 1{2N}e^{-i2\omega\mu}-(-1)^{\omega}e^{-\eta_+\lvert\omega\rvert}\cos(2\omega b)}{1+e^{-2\eta_+\lvert\omega\rvert}},
\quad \omega=\pm1,\pm2,\cdots,\pm\infty,& \\
&\frac12-\frac1{2N},\quad\omega=0.&
\end{aligned}
\right.
\end{eqnarray}
Then the ground state energy is
\begin{eqnarray}\label{Emu}
E(\mu)&=&2N\phi(2a)\sinh\eta \int_{-\frac\pi2}^{\frac\pi2} [\coth(ix+\eta_+-ib-\eta/2) + \coth(ix-\eta_+-ib-\eta/2)\nonumber\\
 &&+\coth(ix+\eta_++ib-\eta/2)+\coth(ix-\eta_++ib-\eta/2)]\rho_2(x)dx\nonumber\\
 &&+\phi(2a)\sinh\eta[\coth(i\mu-ib-\eta/2)+\coth(i\mu+ib-\eta/2) ]+E_0. \nonumber\\
 &=&-4N\frac{\cosh(2\eta_+)-\cos(4b)}{\sinh\eta_+}\sum_{\omega=1}^{\infty}e^{-2\eta_+\omega}\cos^{2}(2b\omega)\tanh(\eta_+\omega)\nonumber\\
&&+2\frac{\cosh(2\eta_+)-\cos(4b)}{\sinh\eta_+}\sum_{\omega=1}^{\infty}(-1)^{\omega}e^{-\eta_+\omega}\cos(2b\omega)\cos(2\mu\omega)\tanh(\eta_+\omega)\nonumber\\
&&+\frac12[\cosh(2\eta_+)-\cos(4b)][\frac{1}{\cosh(\eta_+)+\cos2(\mu+b)}+\frac{1}{\cosh(\eta_+)+\cos2(\mu-b)}]\nonumber\\
&&+\frac{N\cosh\eta_+[\cos^2(2b)-\cosh(2\eta_+)]}{\sinh^2\eta_+}.
\end{eqnarray}

Some remarks are in order. From Eq.(\ref{Emu}), we see that the values of
$E(\mu)$ is dependent on the strength of the boundary string $\mu$.
At the ground state state, $E(\mu)$ should take its minimum. Thus the boundary string at the ground state is fixed.
We find that $E(\mu)$ arrives at its minimum at the point of $\mu=0$ if $b\in(0, \pi/4)$, and at the point of $\mu=-\pi/2$ if $b\in(\pi/4, \pi/2)$.
This conclusion can also be achieved as follows.
If $\eta_+ \to \infty$, many terms in Eq.(\ref{Emu}) tend to zero.
Keeping the order of $e^{-\eta_+}$, the $\mu$-dependent terms in Eq.(\ref{Emu}) can be approximated as
\begin{eqnarray}
&&2\frac{\cosh(2\eta_+)-\cos(4b)}{\sinh\eta_+}\sum_{\omega=1}^{\infty}(-1)^{\omega}e^{-\eta_+\omega}\cos(2b\omega)\cos(2\mu\omega)\tanh(\eta_+\omega)\nonumber\\
&&+\frac12[\cosh(2\eta_+)-\cos(4b)][\frac{1}{\cosh(\eta_+)+\cos2(\mu+b)}+\frac{1}{\cosh(\eta_+)+\cos2(\mu-b)}]\nonumber\\
&&\approx\frac{\cosh(2\eta_+)-\cos(4b)}{\cosh\eta_+}[1-4e^{-\eta_+}\cos(2b)\cos(2\mu)],
\end{eqnarray}
which has the minimum at $\mu=0$ for $b\in(0, \pi/4)$, and at $\mu=-\pi/2$ for $b\in(\pi/4, \pi/2)$.
If $\eta_+$ is finite, we have checked this conclusion numerically and find that it is true.
Therefore, the ground state energy in the regime of $b\in(0, \pi/4)$ (phase I) is
\begin{eqnarray}\label{E2g}
E_{2g}&=& -4N\frac{\cosh(2\eta_+)-\cos(4b)}{\sinh\eta_+}\sum_{\omega=1}^{\infty}e^{-2\eta_+\omega}\cos^{2}(2b\omega)\tanh(\eta_+\omega)\nonumber\\
&&+2\frac{\cosh(2\eta_+)-\cos(4b)}{\sinh\eta_+}\sum_{\omega=1}^{\infty}(-1)^{\omega}e^{-\eta_+\omega}\cos(2b\omega)\tanh(\eta_+\omega)\nonumber\\
&&+\frac{\cosh(2\eta_+)-\cos(4b)}{\cosh(\eta_+)+\cos(2b)}+\frac{N\cosh\eta_+[\cos^2(2b)-\cosh(2\eta_+)]}{\sinh^2\eta_+}.
\end{eqnarray}
The ground state energy in the regime of $b\in(\pi/4, \pi/2)$ (phase II) is
\begin{eqnarray}\label{E3g}
E_{3g}&=& -4N\frac{\cosh(2\eta_+)-\cos(4b)}{\sinh\eta_+}\sum_{\omega=1}^{\infty}e^{-2\eta_+\omega}\cos^{2}(2b\omega)\tanh(\eta_+\omega)\nonumber\\
&&+2\frac{\cosh(2\eta_+)-\cos(4b)}{\sinh\eta_+}\sum_{\omega=1}^{\infty}e^{-\eta_+\omega}\cos(2b\omega)\tanh(\eta_+\omega)\nonumber\\
&&+\frac{\cosh(2\eta_+)-\cos(4b)}{\cosh(\eta_+)-\cos(2b)}+\frac{N\cosh\eta_+[\cos^2(2b)-\cosh(2\eta_+)]}{\sinh^2\eta_+}.
\end{eqnarray}

\begin{figure}[t]
\begin{center}
\includegraphics[width=6cm]{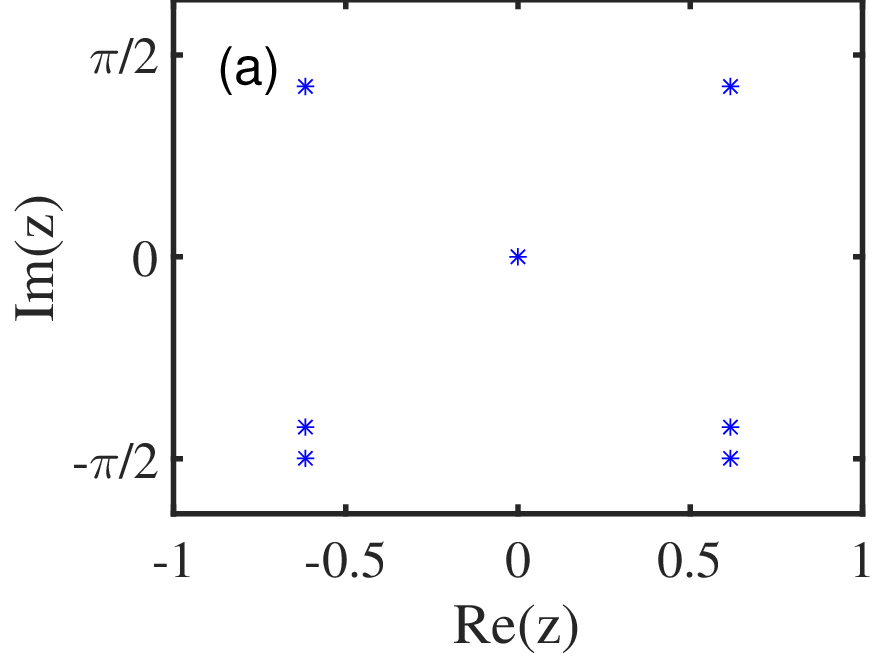}
\includegraphics[width=6cm]{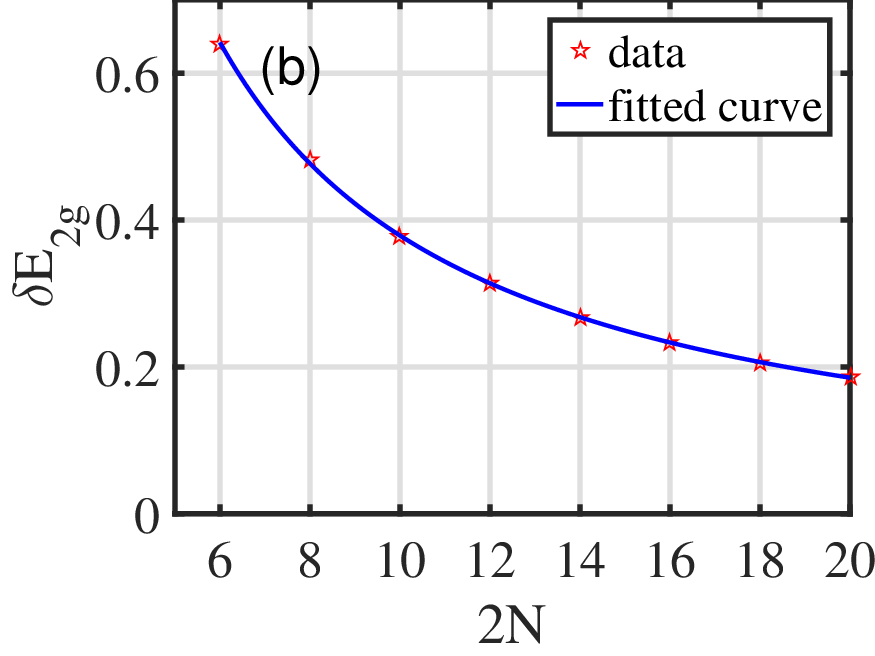}\\
\includegraphics[width=6cm]{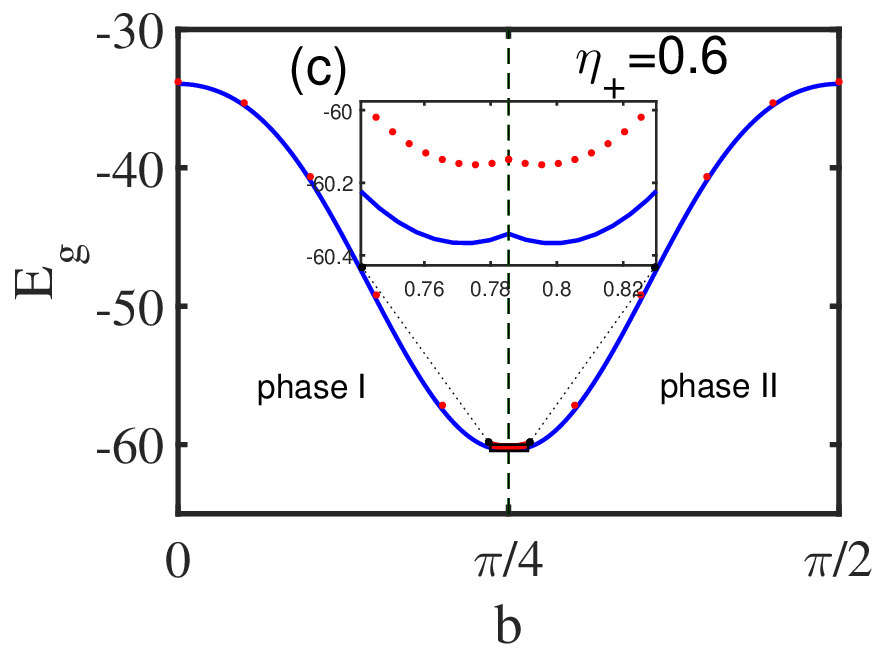}
\includegraphics[width=6cm]{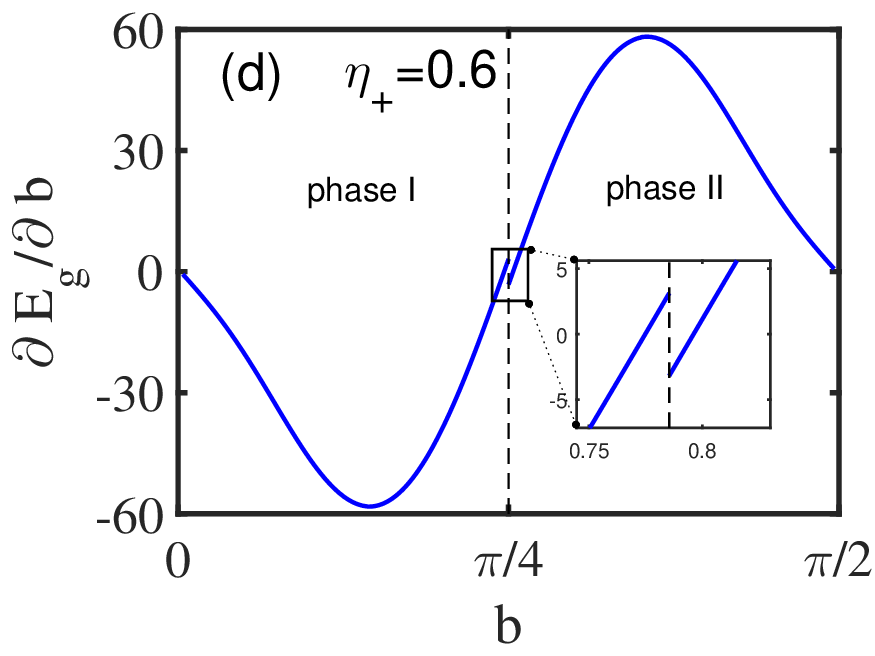}
\caption{The distribution of zero roots $\{z_j\}$ at the ground states with $2N=8$, $\eta_+=0.6$ and $b=0.2$.
 (b) The finite size scaling behavior of $\delta E_{2g}=E_{2g}-E_{2gd}$ with $\eta_{+}=0.6$ and $b=0.2$. Here
 $E_{2g}$ is the analytical result obtained by Eq.(\ref{E2g}) and $E_{2gd}$ is the one calculated by the exact numerical diagonalization.
 The data can be fitted as $\delta E_{2g}=4.076\times(2N)^{-1.032}$.
 Thus $\delta E_{2g}$ tends to zero in the thermodynamic limit.
 (c) The ground state energy versus the interaction $b$, where the blue solid line is the analytical result and red dots are the numerical ones with $2N=18$.
 Comparing the blue solid line and red dots, we see that the finite size scaling effect at $2N=18$ is small.
 (d) The derivative of ground state energy against the interaction $b$.
 At the point of $b=\pi/4$, the derivative is discontinuous.
}\label{fig-E23}
\end{center}
\end{figure}

Now, we check the correctness of analytical expressions (\ref{E2g}) and (\ref{E3g}).
The zero roots distributions with $2N=8$ are shown in Fig.\ref{fig-E23}(a).
The finite size scaling behavior of $\delta E_{2g}=E_{2g}-E_{2gd}$ is shown in Fig.\ref{fig-E23}(b), where
$E_{2g}$ is the analytical result obtained by Eq.(\ref{E2g}) and $E_{2gd}$ is the one calculated by the exact numerical diagonalization.
The data can be fitted as $\delta E_{2g}=4.076\times(2N)^{-1.032}$. Thus $\delta E_{2g}$ tends to zero in the thermodynamic limit.
The numerical results and the analytical ones agree with each other very well.
The ground state energies versus the model parameter $b$ are shown in Fig.\ref{fig-E23}(c),
where the red dots are the numerical data with $2N=18$ and the blue solid lines are the analytical data.
The derivative of the ground state energies versus the interaction $b$ is shown in Fig.\ref{fig-E23}(d).
From it, we see that the ground state energies are continuous and their derivatives are discontinuous.
Thus, there exists a first order quantum phase transition at the critical point $b=\pi/4$.

\subsection{Elementary excitation I}

In the regime of $\eta\in \mathbb{R}+i\pi$, the system has two kinds of elementary excitations. The first kind of excitation is characterized by the root $i\mu$ sliding along the imaginary axis in the interval
$[-i\pi/2, i\pi/2)$ but away from the points of 0 and $-i\pi/2$.
In the regime of $b\in(0, \pi/4)$, the excited energy is
\begin{eqnarray}\label{e2}
e_{2}(\mu)&=& 2\frac{\cosh(2\eta_+)-\cos(4b)}{\sinh\eta_+}\sum_{\omega=1}^{\infty}(-1)^{\omega}e^{-\eta_+\omega}
  [\cos(2\mu\omega)-1]\cos(2b\omega)\tanh(\eta_+\omega)\nonumber\\
&&+\frac12[\cosh(2\eta_+)-\cos(4b)]\left[\frac{1}{\cosh(\eta_+)+\cos2(\mu+b)}\right.\nonumber\\
&&+\left.\frac{1}{\cosh(\eta_+)+\cos2(\mu-b)}-\frac{2}{\cosh(\eta_+)+\cos(2b)}\right].
\end{eqnarray}
The corresponding momentum is
\begin{eqnarray}\label{k2}
k_2(\mu) &=& -i2N\int_{-\frac\pi2}^{\frac\pi2}\ln\frac{\cos(b+x-i\frac{\eta_+}2)\cos(b+x+i\frac{3\eta_+}2)}{\cos(b-x-i\frac{\eta_+}2)\cos(b-x+i\frac{3\eta_+}2)}
   [\rho_2(\mu,x)-\rho_2(\mu=0,x)]dx\nonumber\\
    &&-i\ln\frac{\cos(b+\mu+i\frac{\eta_+}2)}{\cos(b-\mu+i\frac{\eta_+}2)} {~~}mod\,(2\pi)\nonumber\\
    &=&2\sum_{\omega=1}^{\infty}\frac{\sin(2\omega\mu)}\omega\cos(2\omega b)e^{-\eta_+\omega}\tanh(\eta_+\omega)\nonumber\\
    &&+\frac{i}2\bigg[\ln\frac{\cos(b+\mu-i\frac{\eta_+}2)}{\cos(b-\mu-i\frac{\eta_+}2)}
    -\ln\frac{\cos(b+\mu+i\frac{\eta_+}2)}{\cos(b-\mu+i\frac{\eta_+}2)}\bigg] {~~}mod\,(2\pi).
\end{eqnarray}

\begin{figure}[t]
\begin{center}
\includegraphics[width=6cm]{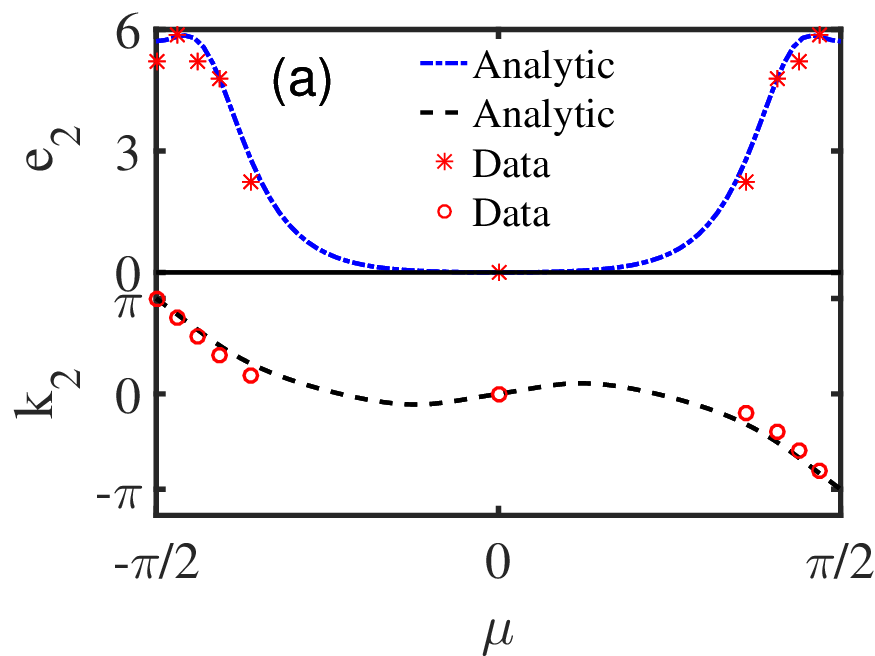}
\includegraphics[width=6cm]{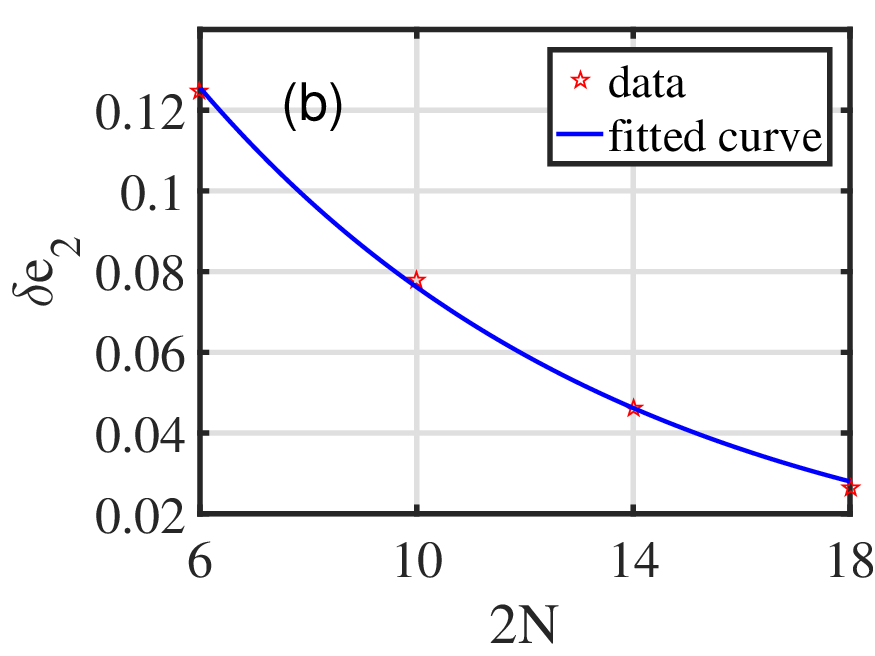}\\
\includegraphics[width=6cm]{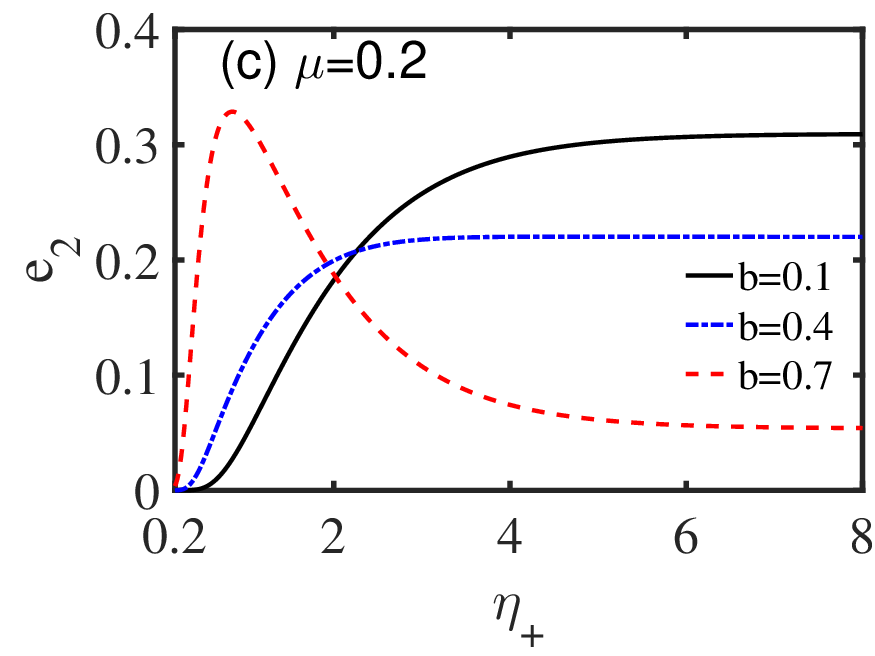}
\includegraphics[width=6cm]{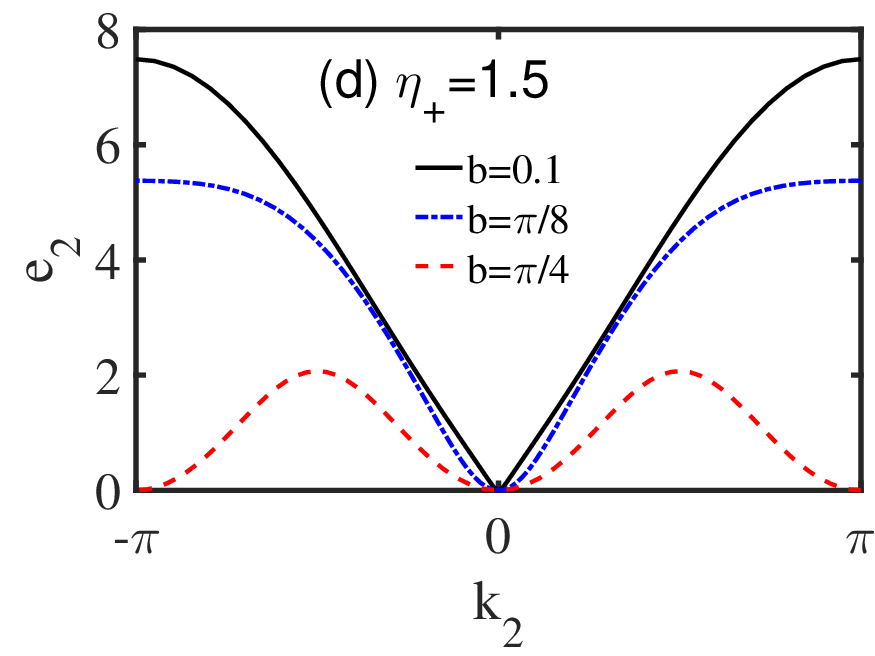}
\caption{The gapless excitation in the antiferromagnetic phase I.
(a) The excited energy $e_2$ and the momentum $k_2$ versus the string strength $\mu$.
The blue solid line is the excited energy calculated from Eq.(\ref{e2}),
the black solid line is the momentum obtained from Eq.(\ref{k2}), and the red circles and starts are
the numerical results computed by using the exact diagonalization with $2N=10$, where $b=0.2$ and $\eta_+=0.6$.
The slight differences are due to the finite size corrections.
(b) The finite size scaling behavior of $\delta e_{2}=e_{2}-e_{2d}$ with $\eta_{+}=1$ and $b=0.75$. Here
 $e_{2}$ is the analytical result obtained by Eq.(\ref{e2}) and $e_{2d}$ is the numerical datum with finite system size.
 The data can be fitted as $\delta e_{2}=0.2658\times e^{-0.125\times 2N}$, which tends to zero in the thermodynamic limit.
(c) The excited energy $e_{2}$ as a function of anisotropic parameter $\eta_+$ with $\mu=0.2$ and $b=0.1, 0.4, 0.7$.
(d) The dispersion relation of the first kind of elementary excitation in the phase I with $\eta_+=1.5$ and $b=0.1, \pi/8, \pi/4$.}\label{fig-ee2}
\end{center}
\end{figure}

Now, we check the corrections of Eqs.(\ref{e2}) and (\ref{k2}).
The excited energy $e_{2}$ and the associated momentum $k_2$ versus the boundary string $\mu$ are demonstrated in Fig.\ref{fig-ee2}(a), where the blue dash-dotted line and the black dash line are the analytical results calculated from Eqs.(\ref{e2}) and (\ref{k2}), and the red stars and circles are the numerical data obtained by
exactly diagonalizing the system with $2N=10$.
From them, we see that the analytic expressions are in good agreement with the numerical results.
The finite size scaling behavior of the energy difference $\delta e_2=e_2-e_{2d}$ is shown in Fig.\ref{fig-ee2}(b),
where $e_2$ is the analytical result obtained by Eq.(\ref{e2}) and $e_{2d}$ is the numerical one.
The data can be fitted as $\delta e_{2}=0.2658\times e^{-0.125\times 2N}$, which tends to zero when $N$ tends to infinite.
Thus the result (\ref{e2}) is correct.
The excited energies with given values of interaction $b$ versus the anisotropic parameter $\eta_+$ are plotted in Fig.\ref{fig-ee2}(c).
We see that if the interaction $b$ is small, the excited energies are increasing with the increasing of $\eta_+$.
While if $b$ is large, the excited energies have a maximum at a suitable value of $\eta_+$.
The dispersion relations between $e_{2}$ and $k_{2}$ with give $b$ are shown in Fig.\ref{fig-ee2}(d).
From it, we see that the excitation is gapless.

\begin{figure}[t]
\begin{center}
\includegraphics[width=6cm]{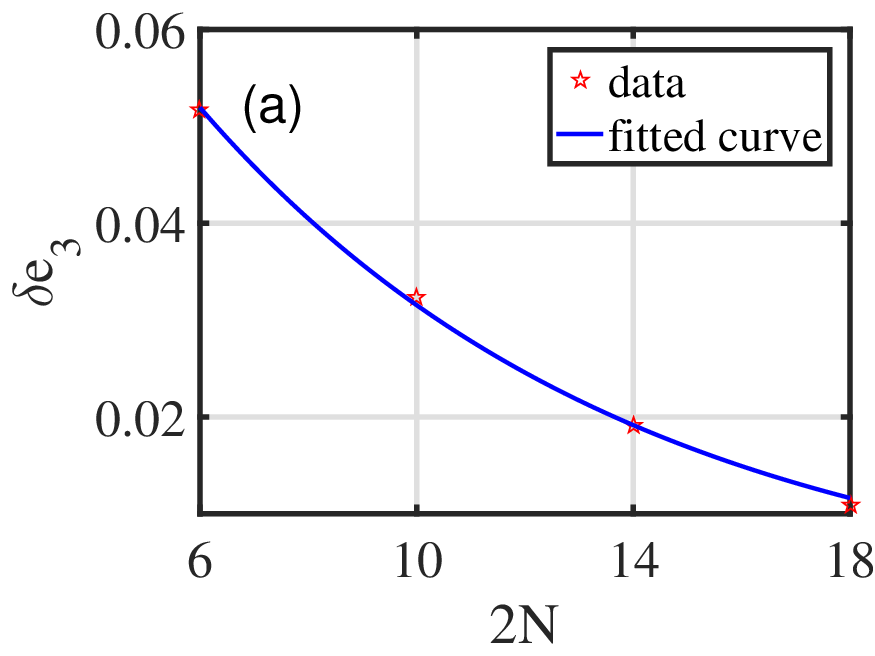}
\includegraphics[width=6cm]{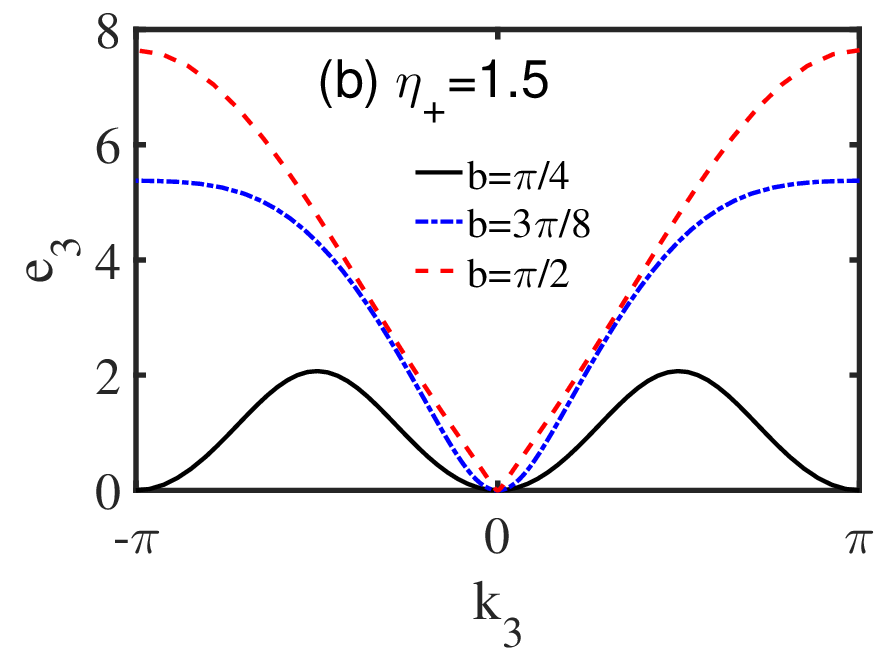}
\caption{ The gapless excitation in the antiferromagnetic phase II.
(a) The finite size scaling behavior of $\delta e_{3}=e_{3}-e_{3d}$ with $\eta_{+}=1$ and $b=0.8$.
Here $e_{3}$ is the analytical result obtained by Eq.(\ref{e3}) and $e_{3d}$ is the numerical one.
 The data can be fitted as $\delta e_{3}=0.1097\times e^{-0.1247\times 2N}$, which is zero in the thermodynamic limit.
(b) The dispersion relation of the first kind of elementary excitation in the phase II with $\eta_+=1.5$ and $b=\pi/4, 3\pi/8, \pi/2$.}
\label{fig-ee3}
\end{center}
\end{figure}

In the regime of $b\in(\pi/4, \pi/2)$, the excited energy is
\begin{eqnarray}\label{e3}
e_{3}(\mu)&=& 2\frac{\cosh(2\eta_+)-\cos(4b)}{\sinh\eta_+}\sum_{\omega=1}^{\infty}e^{-\eta_+\omega}
  [(-1)^{\omega}\cos(2\mu\omega)-1]\cos(2b\omega)\tanh(\eta_+\omega)\nonumber\\
&&+\frac12[\cosh(2\eta_+)-\cos(4b)]\left[\frac{1}{\cosh(\eta_+)+\cos2(\mu+b)}\right.\nonumber\\
&&+\left.\frac{1}{\cosh(\eta_+)+\cos2(\mu-b)}-\frac{2}{\cosh(\eta_+)-\cos(2b)}\right].
\end{eqnarray}
The associated momentum is
\begin{eqnarray}\label{k3}
k_3(\mu) =k_2(\mu)+\pi {~~}mod\,(2\pi).
\end{eqnarray}
The correctness of Eq.(\ref{e3}) is demonstrated by the finite size scaling behavior of $\delta e_3=e_3-e_{3d}$ shown in Fig.\ref{fig-ee3}(a), where $e_2$ is the
excited energy obtained from Eq.(\ref{e3}) and $\delta e_3$ is the numerical one.
We see that the data can be fitted as $\delta e_{3}=0.1097\times e^{-0.1247\times 2N}$, which is zero in the thermodynamic limit.
Based on Eqs.(\ref{e3}) and (\ref{k3}), the dispersion relations in phase II are shown in Fig.\ref{fig-ee3}(b).
We see the excitation is also gapless in this regime.

The physical picture of this kind of elementary excitation is as follows.
From Table \ref{212}, we see that these excited states can be regarded as the superposition of domain walls, which are generated by continuously flipping some spins of anti-ferromagnetic Neel states $|\uparrow\downarrow\cdots\uparrow\downarrow\rangle$ or $|\downarrow\uparrow\cdots\downarrow\uparrow\rangle$.
We should note that the gaps of this kind of states for the system with finite size do not tend to zero in the thermodynamic limit in the regime if $\eta\in \mathbb{R}+i\pi$, thus they are not the nearly degenerate states. If the NN along $x$- and $y$-direction, NNN and chiral three-spin interactions vanish, the model (\ref{Ham1}) reduces to the Ising model.
For the Ising model, these low-lying excited states degenerate to the ground state in the thermodynamic limit.
Therefore, the anisotropic NN, NNN and chiral three-spin interactions separate those excited state away from each other and preserve the finite energy difference even in the thermodynamic limit.

\begin{table}[!h]
\centering
\caption{ The projections $\beta_{ij}=\langle N_j|\psi_i\rangle$ and the error $\bar \delta=\big||\psi_i\rangle-\sum_{j=1}^{8}\beta_{ij}|N_j\rangle\big|$
with $2N=4$, $a=0.2i$ and $\eta_+=2$. Here $\{|\psi_i\rangle, i=1,\cdots,16\}$ are the eigenstates. Among them, the
first two are the ground states, from third to 8-th are the type I low-lying excited states, and the rest ones are the high excited states.
$\{|N_j\rangle=\prod_{k=1}^{j-1}\sigma^x_{(k-1) mod (4)+1}|\uparrow\downarrow\uparrow\downarrow\rangle,j=1,\cdots,8\}$ are the approximate basis vectors of 8-dimensional low-lying states.
The values of $\bar \delta$ at the low-lying states are much smaller than those at the high excited states.}\label{212}
{ \footnotesize
\begin{tabular}{l|llll}
\hline
 $\beta_{ij}$ & $j=1$ &  $j=2$ &   $j=3$& $j=4$ \\ \hline
$i=1$ & $0.4972-0.0000i$ & $-0.0000+0.0000i$ & $0.4972$ & $-0.0000+0.0000i$ \\
$i=2$ & $0.2300+0.1375i$ & $0.4188+0.0000i$ & $0.2300+0.1375i$ & $0.4188$ \\
$i=3$ & $-0.1834-0.2577i$ & $-0.4905-0.0000i$ & $0.2971-0.2132i$ & $0.1231-0.0456i$ \\
$i=4$ & $-0.2413-0.1058i$ & $-0.2207-0.3389i$ & $-0.4955+0.0000i$ & $0.1003-0.0541i$ \\
$i=5$ & $0.0703+0.0084i$ & $-0.0999-0.3037i$ & $0.0599+0.0260i$ & $-0.6164-0.0000i$ \\
$i=6$ & $-0.1425-0.1667i$ & $0.2350+0.3183i$ & $-0.0877+0.0745i$ & $0.5231$ \\
$i=7$ & $-0.4927+0.0000i$ & $0.0000-0.0000i$ & $0.4927$ & $-0.0000+0.0000i$ \\
$i=8$ & $0.0450+0.0186i$ & $0.4903$ & $-0.0450-0.0186i$ & $-0.4903-0.0000i$ \\
$i=9$& $-0.0461-0.0264i$ & $-0.0000-0.0000i$ & $-0.0461-0.0264i$ & $0.0000+0.0000i$ \\
$i=10$ & $0.0003-0.0002i$ & $-0.0461+0.0264i$ & $0.0003-0.0002i$ & $-0.0461+0.0264i$ \\
$i=11$& $-0.0120+0.0007i$ & $0.0020-0.0048i$ & $-0.0835-0.0366i$ & $0.0014-0.0002i$ \\
$i=12$ & $0.0797+0.0350i$ & $-0.0159+0.0149i$ & $-0.0087-0.0102i$ & $-0.0156-0.0030i$ \\
$i=13$ & $0.0102+0.0042i$ & $-0.0788+0.0346i$ & $-0.0023-0.0046i$ & $0.0299+0.0064i$ \\
$i=14$& $0.0688+0.0302i$ & $0.0270-0.0114i$ & $-0.0065-0.0070i$ & $0.0415+0.0130i$ \\
$i=15$& $-0.0811-0.0257i$ & $-0.0000-0.0000i$ & $0.0811+0.0257i$ & $0.0000+0.0000i$ \\
$i=16$ & $-0.0023-0.0001i$ & $-0.0811+0.0257i$ & $0.0023+0.0001i$ & $0.0811-0.0257i$ \\
\hline
\end{tabular}
\begin{tabular}{llll|l}
\hline
 $j=5$ &$j=6$  &  $j=7$&  $j=8$ & $\bar \delta$\\ \hline
  $0.4972-0.0000i$ & $-0.0000+0.0000i$ & $0.4972+0.0000i$ & $-0.0000+0.0000i$ &  $0.1062$ \\
 $0.2300+0.1375i$ & $0.4188+0.0000i$ & $0.2300+0.1375i$ & $0.4188+0.0000i$ &  $0.1062$ \\
 $0.1834+0.2577i$ & $0.4905$ & $-0.2971+0.2132i$ & $-0.1231+0.0456i$ &  $0.1302$ \\
 $0.2413+0.1058i$ & $0.2207+0.3389i$ & $0.4955$ & $-0.1003+0.0541i$ &  $0.1302$ \\
 $-0.0703-0.0084i$ & $0.0999+0.3037i$ & $-0.0599-0.0260i$ & $0.6164$ &  $0.1302$ \\
 $0.1425+0.1667i$ & $-0.2350-0.3183i$ & $0.0877-0.0745i$ & $-0.5231+0.0000i$ &  $0.1302$ \\
 $-0.4927+0.0000i$ & $-0.0000-0.0000i$ & $0.4927$ & $-0.0000+0.0000i$ &  $0.1702$ \\
 $0.0450+0.0186i$ & $0.4903+0.0000i$ & $-0.0450-0.0186i$ & $-0.4903-0.0000i$ &  $0.1702$ \\
 $-0.0461-0.0264i$ & $-0.0000-0.0000i$ & $-0.0461-0.0264i$ & $0.0000+0.0000i$ &  $0.9943$ \\
 $0.0003-0.0002i$ & $-0.0461+0.0264i$ & $0.0003-0.0002i$ & $-0.0461+0.0264i$ &  $0.9943$ \\
 $0.0120-0.0007i$ & $-0.0020+0.0048i$ & $0.0835+0.0366i$ & $-0.0014+0.0002i$ &  $0.9915$ \\
 $-0.0797-0.0350i$ & $0.0159-0.0149i$ & $0.0087+0.0102i$ & $0.0156+0.0030i$ &  $0.9915$ \\
 $-0.0102-0.0042i$ & $0.0788-0.0346i$ & $0.0023+0.0046i$ & $-0.0299-0.0064i$ &  $0.9915$ \\
 $-0.0688-0.0302i$ & $-0.0270+0.0114i$ & $0.0065+0.0070i$ & $-0.0415-0.0130i$ &  $0.9915$ \\
 $-0.0811-0.0257i$ & $-0.0000-0.0000i$ & $0.0811+0.0257i$ & $0.0000+0.0000i$ &  $0.9854$ \\
 $-0.0023-0.0001i$ & $-0.0811+0.0257i$ & $0.0023+0.0001i$ & $0.0811-0.0257i$ &  $0.9854$ \\
\hline
\end{tabular}}
\end{table}

The competing of the anisotropic NN, NNN and chiral three-spin interactions will change these excited energies and induce the quantum phase transition.
In the phase I, the NN interactions along the $x$- and $y$-directions are ferromagnetic.
With the increasing of interaction parameter $b$, the energies for modes of momentum $\pi$ or $-\pi$ are decreasing and eventually tend to zero, please see Fig.\ref{fig-ee2}(d).
If the value of $b$ is larger than the critical point $\pi/4$, the NN interactions along the $x$- and $y$-directions become antiferromagnetic,
and the contribution of chiral three-spin interactions along the $z$-direction also changes.
The energy of $\pi$ mode is lower than the ground state energy in phase I, and the corresponding state becomes the new ground state in phase II.
From Fig.\ref{fig-ee2}(a), we also see that the momentum $\pi$ corresponds to $\mu=-\pi/2$, which determines the ground state in phase II.
This result agrees with that given by Eq.(\ref{E3g}).
Then we demonstrate that there exists a quantum phase transition from the phase I to the phase II.
Thus the competition have significantly influenced this kind of low-lying gapless excitations and can induce the phase transition.

\subsection{Elementary excitation II}

\begin{figure}[t]
\begin{center}
\includegraphics[width=6cm]{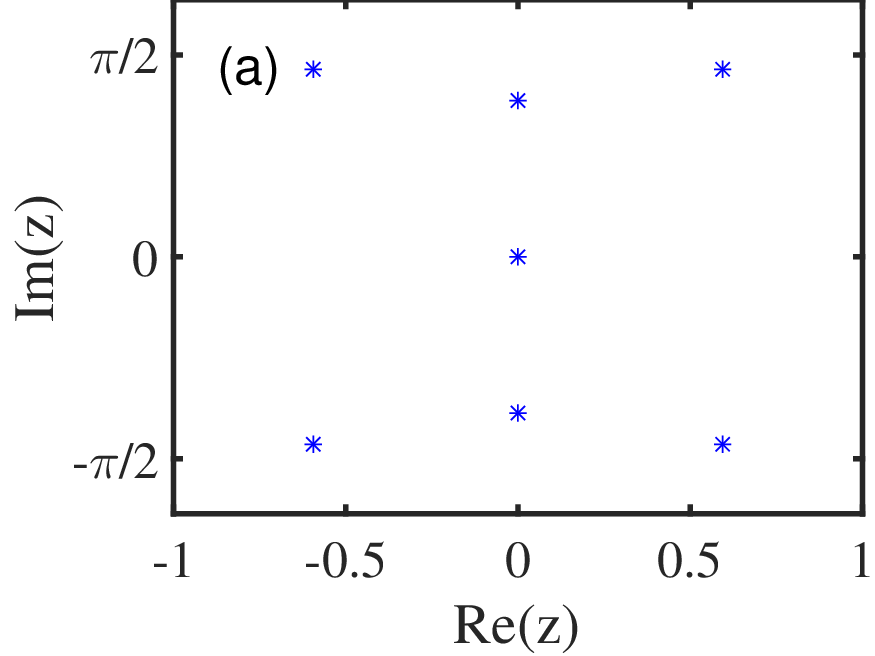}
\includegraphics[width=6cm]{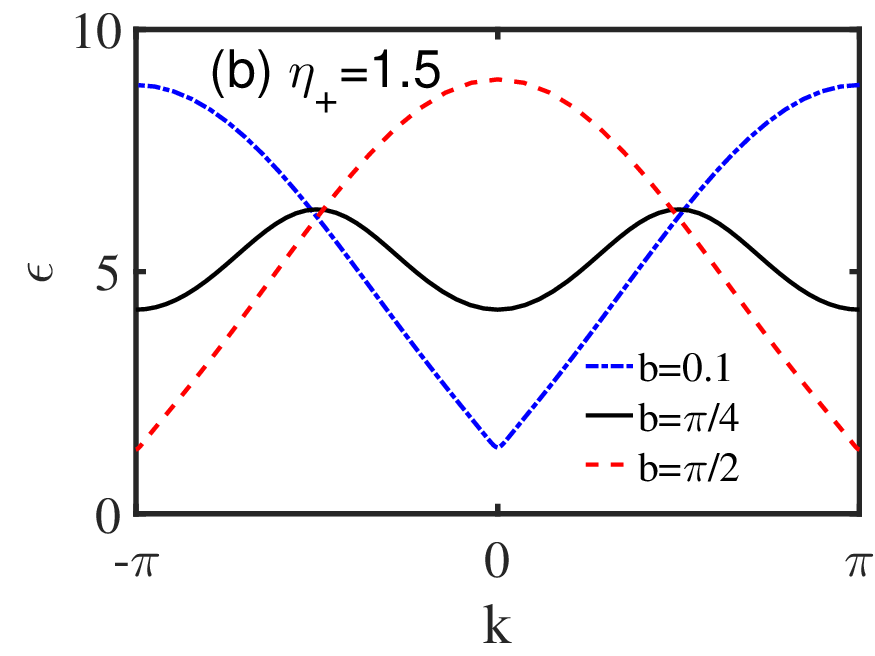}
\caption{ The gapped spinon excitations.
(a) The pattern of zero roots with $2N=8$, $\eta_+=0.6$ and $b=0.2$.
(b) The dispersion relations with $\eta_+=1.5$ and $b=0.1, \pi/4, \pi/2$.}\label{fig-ee4}
\end{center}
\end{figure}

The second kind of elementary excitations is quantified by a conjugate pair of $z$-roots turning into two imaginary ones, i.e.,
\begin{eqnarray}\label{zroot3}
&&z_{2j-1}=ix_{2j-1}+\eta_++o(e^{-\delta N}),\quad z_{2j}=ix_{2j}-\eta_++o(e^{-\delta N}),\quad j=1,\cdots,N-2,\nonumber\\
&&z_{2N-3}=i\mu_1+o(e^{-\delta N}),\quad z_{2N-2}=i\mu_2+o(e^{-\delta N}),\nonumber\\
&&z_{2N-1}=i\mu,
\end{eqnarray}
where $x_{j}$, $\mu_{1}$, $\mu_{2}$ and $\mu$ are all real.
The root patterns of such excitations for the phases I are shown in Fig.\ref{fig-ee4}(a).
Using the similar procedure mentioned above, we obtain the density of zero roots in the thermodynamic limit should satisfy the integral equation
\begin{eqnarray}
&&-c_1(\phi-\mu)-c_1(\phi-\mu_1)-c_1(\phi-\mu_2)-2N\int_{-\frac\pi2}^{\frac\pi2} [c_1(\phi-x)+c_3(\phi-x)]\rho_2(x)dx\nonumber\\ &&=2N\int_{-\frac\pi2}^{\frac\pi2} b_2(\phi-x)\sigma(x)dx.
\end{eqnarray}
With the help of Fourier transformation, we obtain the solution as
\begin{eqnarray}
\tilde{\rho}_3(\omega)=\left\{
\begin{aligned}
&-\frac{\frac 1{2N}(e^{-i2\omega\mu}+e^{-i2\omega\mu_1}+e^{-i2\omega\mu_2})-(-1)^{\omega}e^{-\eta_+\lvert\omega\rvert}\cos(2\omega b)}{1+e^{-2\eta_+\lvert\omega\rvert}},
\;\; \omega=\pm1,\cdots,& \\
&\frac12-\frac1{N},\quad\omega=0.&
\end{aligned}
\right.\nonumber
\end{eqnarray}
We find that the excited energies in phases I and II have the same expression
\begin{eqnarray}\label{e4}
e_{4}(\mu_1,\mu_2)&=&E(\mu_1,\mu_2,\mu=0)-E(\mu=0)=E(\mu_1,\mu_2,\mu=-\frac\pi2)-E(\mu=-\frac\pi2)\nonumber\\
&=&\epsilon(\mu_1)+\epsilon(\mu_2),
\end{eqnarray}
where
\begin{eqnarray}
\epsilon(\mu)&=& 2\frac{\cosh(2\eta_+)-\cos(4b)}{\sinh\eta_+}\sum_{\omega=1}^{\infty}(-1)^{k}e^{-\eta_+\omega}
  \cos(2\mu\omega)\cos(2b\omega)\tanh(\eta_+\omega)\nonumber\\
&&+\frac12[\cosh(2\eta_+)-\cos(4b)]\left[\frac{1}{\cosh(\eta_+)+\cos2(\mu+b)}\right.\nonumber\\
&&+\left.\frac{1}{\cosh(\eta_+)+\cos2(\mu-b)}\right].
\end{eqnarray}
The corresponding momentum reads
\begin{eqnarray}\label{k4}
k_4(\mu_1,\mu_2) &=& -i2N\int_{-\frac\pi2}^{\frac\pi2}\ln\frac{\cos(b+x-i\frac{\eta_+}2)\cos(b+x+i\frac{3\eta_+}2)}{\cos(b-x-i\frac{\eta_+}2)\cos(b-x+i\frac{3\eta_+}2)}
   [\rho_3(\mu_2,\mu_1,\mu,x)-\rho_2(\mu,x)]dx\nonumber\\
    &&-i\ln\frac{\cos(b+\mu_1+i\frac{\eta_+}2)}{\cos(b-\mu_1+i\frac{\eta_+}2)}-i\ln\frac{\cos(b+\mu_2+i\frac{\eta_+}2)}{\cos(b-\mu_2+i\frac{\eta_+}2)} {~~}mod\,(2\pi)\nonumber\\
    &=&k_2(\mu_1)+k_2(\mu_2) {~~}mod\,(2\pi).
\end{eqnarray}

From Eqs.(\ref{e4}) and (\ref{k4}), we see that both the energy and the momentum of this kind of elementary excitations depend on two free parameters.
They are the typical spinon excitations. The dispersion relations with given $b$ are shown in Fig.\ref{fig-ee4}(b).
We see that there always exists an energy gap and this kind of excitations is gapped.

\section{Conclusions}

In this paper, we have studied the exact physical properties of an integrable antiperiodic $J_1-J_2$ spin chain that includes the NN, NNN and chiral three-spin interactions in the thermodynamic limit. With the help of the inhomogeneous $T-Q$ relation, we obtain the zero roots distributions of the transfer matrix focusing on the interaction parameter $a$ is imaginary and $\eta$ is real or $\eta\in \mathbb{R}+i\pi$. Based on the root patterns, we calculate the ground state energies, elementary excitations and dispersion relations.
We also discuss the nearly degenerate states in the ferromagnetic regime with $\eta\in \mathbb{R}$ and the quantum phase transition
in the antiferromagnetic regime with $\eta\in \mathbb{R}+i\pi$.
We demonstrate the competing of NN, NNN and chiral three-spin interactions can induce many interesting phenomena.

\section*{Acknowledgments}

The financial supports from the National Key R\&D Program of China (Grants Nos. 2021YFA1402104, 2021YFA0718300, 2021YFA1400900 and 2021YFA1400243),
the National Natural Science Foundation of China (Grant Nos. 61835013, 12074178, 12074410, 11934015,
11975183, 12147160, 11774150 and 12234012), the Strategic
Priority Research Program of the Chinese Academy of Sciences (Grant Nos. XDB33000000),
Space Application System of China Manned Space Program
and the fellowship of China Postdoctoral Science Foundation (Grant No. 2020M680724) are gratefully acknowledged.

\end{document}